\documentclass[preprint,12pt,3p,onecolumn]{elsarticle}
\biboptions{sort&compress}

\usepackage[utf8x]{inputenc}
\usepackage{amsmath}
\usepackage{amssymb}
\usepackage{color}
\usepackage{setspace}
\usepackage{graphicx}
\usepackage{subfigure}
\usepackage{geometry}
\usepackage[format=hang,font=small,labelfont=bf]{caption}
\usepackage{multirow}
\usepackage{multicol}
\usepackage{algorithm}
\usepackage{algpseudocode}
\usepackage{fixltx2e}
\usepackage{url}
\usepackage{placeins}
\usepackage{lineno}

\definecolor{dkgreen}{rgb}{0,0.6,0}
\definecolor{dgreen}{rgb}{0,0.6,0.2}
\definecolor{gray}{rgb}{0.5,0.5,0.5}
\definecolor{dark}{rgb}{0.4,0.4,0.4}
\definecolor{silver}{rgb}{0.85,0.85,0.85}
\definecolor{mauve}{rgb}{0.58,0,0.82}

\MakeRobust{\Call}
\newcommand{\ongpu}[1]{\colorbox{silver}{#1}}
\algnewcommand{\LComment}[1]{\State \(\triangleright\) {\color{black} \textbf{#1}}}

\usepackage{listings}
\journal{Computer Physics Communications}

\begin{document}


\begin{frontmatter}

\title{Accelerating Dissipative Particle Dynamics Simulations on GPUs: Algorithms, Numerics and Applications}

\author[brown]{Yu-Hang Tang}
\ead{yuhang\_tang@brown.edu}

\author[brown]{George Em Karniadakis\corref{cor1}\fnref{fn1}}
\ead{george\_karniadakis@brown.edu}

\cortext[cor1]{Corresponding author}
\fntext[fn1]{Tel: +1-401-863-1217}
\address[brown]{Division of Applied Mathematics, Brown University, Providence, Rhode Island, USA}

\begin{abstract}

\noindent We present a scalable dissipative particle dynamics simulation code, fully implemented on the Graphics Processing Units (GPUs) using a hybrid CUDA/MPI programming model, which achieves 10-30 times speedup on a single GPU over 16 CPU cores and almost linear weak scaling across a thousand nodes. A unified framework is developed within which the efficient generation of the neighbor list and maintaining particle data locality are addressed. Our algorithm generates strictly ordered neighbor lists in parallel, while the construction is deterministic and makes no use of atomic operations or sorting. Such neighbor list leads to optimal data loading efficiency when combined with a two-level particle reordering scheme. A faster \textit{in situ} generation scheme for Gaussian random numbers is proposed using precomputed binary signatures. We designed custom transcendental functions that are fast and accurate for evaluating the pairwise interaction. The correctness and accuracy of the code is verified through a set of test cases simulating Poiseuille flow and spontaneous vesicle formation. Computer benchmarks demonstrate the speedup of our implementation over the CPU implementation as well as strong and weak scalability. A large-scale simulation of spontaneous vesicle formation consisting of 128 million particles was conducted to further illustrate the practicality of our code in real-world applications.

\end{abstract}

\begin{keyword}

DPD \sep CUDA \sep LAMMPS \sep spontaneous vesicle formation

\end{keyword}

\end{frontmatter}


\section{Introduction}

Particle-based simulation is a powerful tool in studying microscopic stochastic systems and has been popularized by the ever growing computing power provided by each new generation of hardware. Among the spectrum of parallel processors, the general purpose graphics processing unit (GPGPU) proves to be a particularly good fit for such simulation due to the massively parallel nature shared by inter-particle interaction evaluation and image rendering \cite{Anderson20085342,JCC:JCC20829,Liu2008634,JCC:JCC21209}. Specifically, the Compute Unified Device Architecture (CUDA) has provided a parallel programming model that could harness the processing power of the GPGPUs, which is tens of times more than the contemporary CPUs \cite{cuda2008experience, Nickolls2008scalable}. There is currently an array of molecular dynamics simulation applications that integrate the capability to use CUDA for part or all of the computation, such as AMBER, DL\_POLY, HOOMD-BLUE, LAMMPS, NAMD and GROMACS \cite{Amber_GPU, Smith1996136, Anderson20085342, LAMMPS1995, Kale1999283, Pronk01042013}.

The combination of fast hardware and highly optimized software package enables large-scale simulations of atomistic systems. For example, a protein folding simulation of 64 million particles using an all-atom resolution has been successfully carried out using NAMD for a total integration time of 100 ns \cite{zhao2013mature}. Nevertheless, the classical all-atom molecular dynamics (AAMD) method is still prohibitively expensive for simulating atomistic systems at a length and time scale that is comparable to that in experiment. To bridge the gap between the microscopic time/length scale and the macroscopic ones, coarse-grained simulation techniques such as coarse-grained molecular dynamics (CGMD), dissipative particle dynamics (DPD) and smooth particle hydrodynamics (SPH) have been proposed in which a group of atoms in the AAMD model is represented by a single coarse-grained particle to reduce the computational complexity \cite{Hoogerbrugge1992DPD, groot1997dissipative, espanol1995dpd, Monaghan2012sph, Ottinger2013Generic}.

Among the various coarse-grained models, DPD has been our particular interest since it accurately reproduces the equilibrium and dynamic properties of systems in the area of biophysics, soft matter and fluid dynamics \cite{li2009vesicle, Peng13082013, Fedosov19072011}. Unfortunately, little effort has been directed to the efficient implementation of DPD on the GPUs. Despite the intrinsic similarity in the governing principles, the DPD systems differ from their atomistic counterparts in that:
\begin{itemize}
  \item A more complex functional form is used, \textit{i.e.} a dissipative force, which involves the relative velocity in addition to the relative position of the interacting particles.
  \item A pairwise random force is essential to compensate for the reduced degrees of freedom in order to correctly reproduce the dynamics of the coarse-grained system.
  \item The system is sparse, which results in a much lower average neighbor count for the particles and makes it harder to achieve optimal performance on the GPU's SIMD architecture.
\end{itemize}

So far, Goga \textit{et al.} have presented a GPU parallelization of the SD and DPD types in the GROMACS tools \cite{goga2012gpusd}, with which they observed a performance boost of 70\% on the GPU over a single processor. Phillips \textit{et al.} proposed a hash-function based pseudo-random number generator for the evaluation of the stochastic term \cite{TEA_DPD}. Wu \textit{et al.} reported a GPU implementation of DPD with parallel neighbor list updating \cite{wu2011gpu}. Wang \textit{et al.} reported a multi-GPU implementation, which achieves a speedup of 90x on three GPUs over a single-threaded CPU implementation from Material Studio \cite{Wang20132454}. However, no scalable implementation that runs across nodes has been reported so far.

In this paper, we demonstrate a scalable implementation of the DPD simulation on the latest Kepler GPGPU architecture using a CUDA-MPI hybrid programming model targeting optimal performance within a GPU and also across nodes. The paper is organized as follows: Section~\ref{sec:dpd} presents a brief review of the functional forms of DPD. Section~\ref{sec:overview} presents an overview of the program flow, while detailed algorithms are given in Section~\ref{sec:sortingbinningreordering}-\ref{sec:comm}. Benchmarks and verifications of the code simulating simple flow and spontaneous vesicle formation are given in Section~\ref{sec:benchmark}. In Section~\ref{sec:conslusion} we summarize our work.

\section{Functional Forms of DPD}
\label{sec:dpd}

In the DPD simulation, the movement of particles are governed by the Newton's equation of motion as \cite{groot1997dissipative}

\begin{align}
\frac{d\textbf{r}}{d\textbf{t}} = \textbf{v},\;\;\; \frac{d\textbf{v}}{d\textbf{t}} = \textbf{a} = \frac{\textbf{f}}{m}
\end{align}

The force $\textbf{f}$ acting on each particle consists of three pairwise additive parts: a conservative component, a dissipative component and a random component, \textit{i.e.}
\begin{align}
\textbf{f}_i = \sum_{i\neq j} \textbf{F}_{ij} = \sum_{i\neq j} ( \textbf{F}_{ij}^C + \textbf{F}_{ij}^D + \textbf{F}_{ij}^R )
\end{align}
given by
\begin{align}
\textbf{F}_{ij}^C &=       a_{ij} w_C(\textbf{r}_{ij})  \textbf{e}_{ij}                                        \\
\textbf{F}_{ij}^D &= -\gamma_{ij} w_D(\textbf{r}_{ij}) (\textbf{e}_{ij} \cdot \textbf{v}_{ij}) \textbf{e}_{ij} \\
\textbf{F}_{ij}^R &=  \sigma_{ij} w_R(\textbf{r}_{ij})  \xi_{ij} {\delta t}^{-\frac{1}{2}} \textbf{e}_{ij}     
\end{align}
if $ |\textbf{r}_{ij}| \leq r_c $, where $r_c$ is the cutoff distance, and
\begin{align}
\textbf{F}_{ij}^C = \textbf{F}_{ij}^D &= \textbf{F}_{ij}^R = 0,\;\;\; |\textbf{r}_{ij}| > r_c.
\end{align}

One of the two weight functions $w_D$ and $w_R$ can be chosen arbitrarily, while the other is then fixed through the fluctuation-dissipation constraint \cite{Espanol1995}

\begin{align}
w_D(\textbf{r}_{ij}) = w_R^2(\textbf{r}_{ij})
\end{align}
In our implementation, $w_R$ is chosen to be a power of $w_C$ as
\begin{align}
w_R(\textbf{r}_{ij}) = w_C^s(\textbf{r}_{ij}) = (1-\frac{|\textbf{r}_{ij}|}{r_c})^s
\end{align}
The exponent $s$ can be chosen arbitrarily, making it a convenient tuning parameter for reproducing the viscosity of a wide range of fluids. The coefficients $\sigma_{ij}$ and $\gamma_{ij}$ are related to each other by
\begin{align}
\sigma_{ij}^2 = 2 \gamma_{ij} k_B T
\end{align}
as also dictated by the fluctuation-dissipation theorem.

\section{Algorithms and Implementation}
\label{sec:alg}

\subsection{Program Summary}
\label{sec:overview}

We start with LAMMPS as the base code for the CPU portion of the program \cite{LAMMPS1995}. The GPU code is written in CUDA C due to the high level of availability and maturity. The basic control flow is shown in Algorithm~\ref{ALG_LAMMPS} where the highlighted text indicates parts that are accelerated by the GPU. The code essentially runs in a time-stepping loop where tasks such as trajectory integration, communication, force evaluation, statistical analysis and data I/O are executed in order according to the Velocity Verlet algorithm \cite{swope1982vv}. A spatial decomposition scheme is used for parallel execution over MPI, while stray particles that run across domain boundaries are exchanged each time the neighbor list is rebuilt.

\begin{algorithm}
\caption{Basic program control flow. The highlighted text indicates portions of the computation that involve the GPU.}
\label{ALG_LAMMPS}
\begin{algorithmic}[1]
\LComment{setup}
\State setup compute domain, determine particle ownership
\State \ongpu{CPU $\xrightarrow{\mathrm{data}}$ GPU}
\State \ongpu{obtain ghost particles information from neighboring processors}
\State \ongpu{build neighbor list}
\State \ongpu{compute forces}
\LComment{main loop}
\For{n steps} 
  \State \ongpu{time integration: phase 1}
  \If{time to rebuild neighbor list}
    \State \ongpu{GPU $\xrightarrow{\mathrm{data}}$ CPU}
    \State exchange stray particles
    \State \ongpu{CPU $\xrightarrow{\mathrm{data}}$ GPU}
    \State \ongpu{obtain ghost particles information from neighboring processors}
    \State \ongpu{rebuild neighbor list}
  \Else
    \State \ongpu{exchange ghost particle information}
  \EndIf
  \State \ongpu{compute forces}
  \State \ongpu{time integration: phase 2}
  \State \ongpu{calculate statistics}
  \If {on demand} \State display runtime info, dump trajectory, write restart file, etc. \EndIf
\EndFor
\end{algorithmic}
\end{algorithm}

\begin{figure}[htbp]
\centering
\includegraphics[width=3.0in]{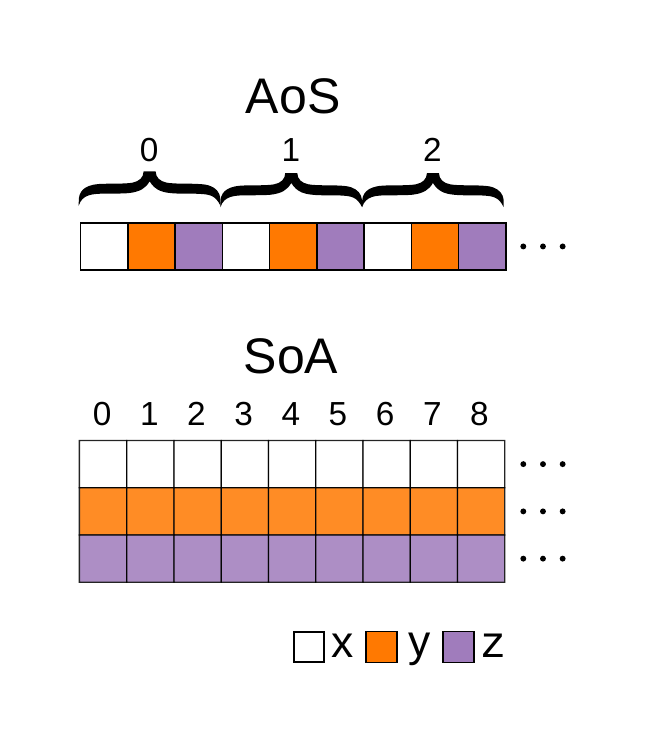}
\caption{In the array of structure (AoS) layout, the coordinate vector for each particle is placed consecutively, whereas any specific component of the vector are separated by the other ones. In the structure of array (SoA) data layout, components are stored consecutively.}
\label{fig:AoS_SoA}
\end{figure}

LAMMPS uses an Array of Structures (AoS) layout to store vector-valued particle properties such as coordinate, velocity and force as illustrated in Figure~\ref{fig:AoS_SoA}. Such layout results in strided access on the GPU, \textit{i.e.} threads with consecutive thread IDs will access memory locations that are separated by a stride larger than one, which reduces the effective memory bandwidth. On the other hand, too much code modification is required if we were to change a data layout which is used essentially everywhere in the original program. As a compromise we used a pair of interleave/deinterleave kernels to convert data between the AoS and the SoA layouts at runtime. The kernels are carefully designed to achieve high memory bandwidth efficiency as summarized in Table~\ref{table:interleave_bandwidth}.

\subsection{Sorting, Binning and Reordering}
\label{sec:sortingbinningreordering}

\subsubsection{Sorting}

\begin{figure}[htbp]
\centering
\includegraphics[width=4.0in]{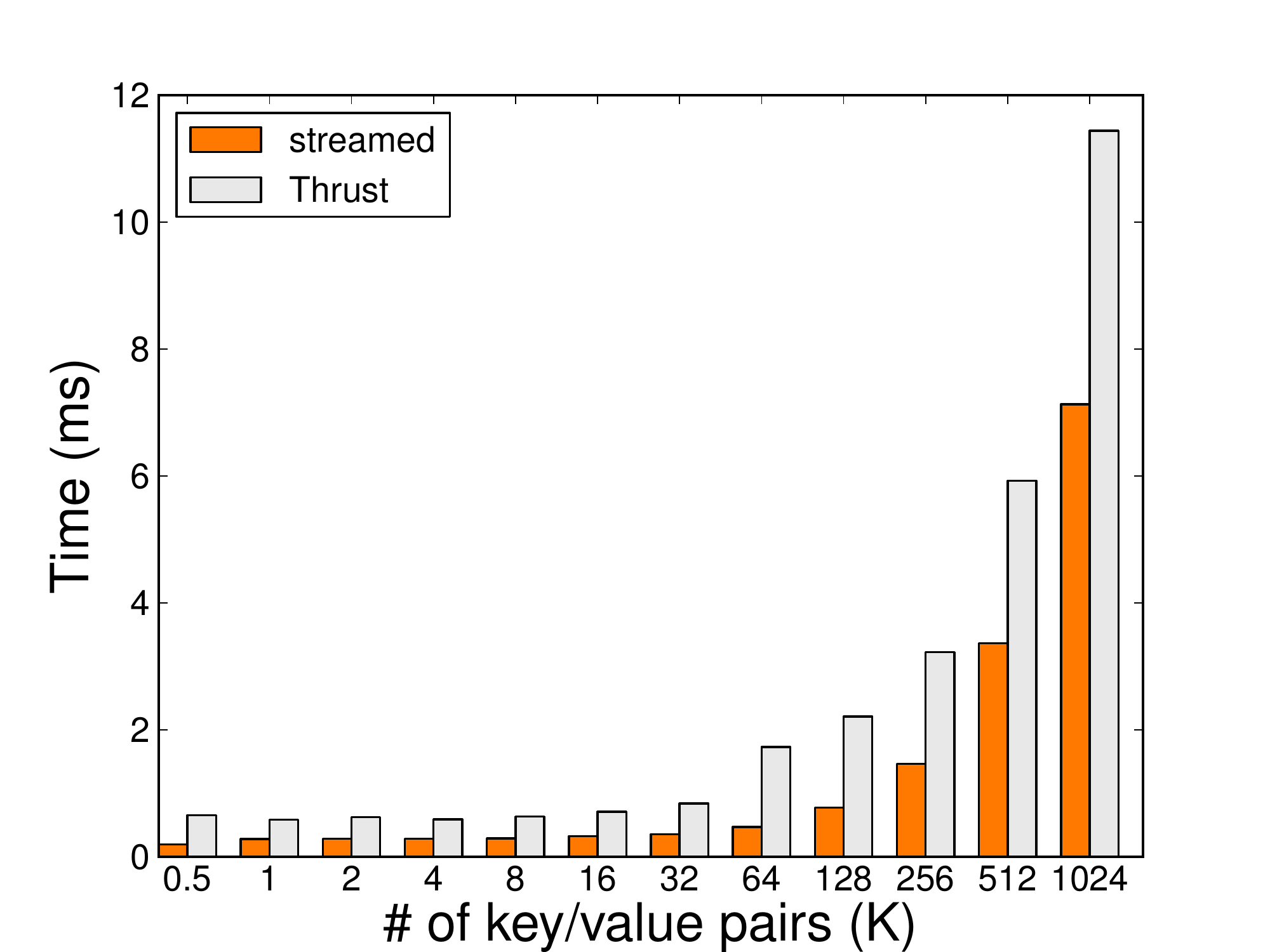}
\caption{Sorting time: our streamed radix sort \textit{vs.} Thrust.}
\label{fig:radix}
\end{figure}

The parallel sorting primitive plays a fundamental role in our code for reordering particles and building the cell list. Several existing algorithm packages such as Thrust and CUDA Data Parallel Primitives Library (CUDPP) provide high-level interfaces for this type of workload (CUDPP actually uses Thrust::sort as its backend) \cite{CUDPP_radix, Thrust}. However, neither package supports the CUDA streams. In other words, they can only launch kernels on the default stream $0$. In our code, streaming is essential for achieving high GPU computing efficiency as it helps to hide communication and kernel launching latency. Moreover, our benchmark reveals that these packages are better optimized for large arrays with millions or billions of key/value pairs, while in practical DPD simulations the particle number per GPU does not exceed a few millions. Therefore, we implemented our own radix sort that is optimized specifically for smaller arrays. As shown in Algorithm~\ref{ALG_RADIX}, each successive invocation of three kernels \texttt{gpuRadixHistogram}, \texttt{gpuRadixPrefixSum} and \texttt{gpuRadixPermute} completes one pass of the sorting algorithm, while 4 bit per pass was determined to be the optimal radix width through experimentation. The scan primitive by Sengupta \textit{et al.} combined with the newly-introduced warp shuffle instructions is used to carry out the prefix summation \cite{scan_primitive}. The benchmark presented in Figure~\ref{fig:radix} compares the performance of our implementation versus Thrust over a wide range of array sizes.

\begin{algorithm}
\caption{Pseudo code for the radix sort template function.}
\label{ALG_RADIX}
\begin{algorithmic}[1]
\Function{RadixSort<RADIX>}{KeyArray, ValArray, BitLength, TargetStream}
  \State BufferIn  $\gets$ \{KeyArray, ValArray\}
  \State BufferOut $\gets$ InternalBuffer
  \For{ $bit = 0$ \textbf{to} BitLength \textbf{by} \Call{Log2}{RADIX} }
    \State \Call{gpuRadixHistogram}{BufferIn,BufferOut,PrefixBuffer,bit} on TargetStream
    \State \Call{gpuRadixPrefixSum}{PrefixBuffer} on TargetStream
    \State \Call{gpuRadixPermute}{BufferIn,BufferOut,PrefixBuffer,bit} on TargetStream
    \State \Call{SwapPointer}{BufferIn, BufferOut}
  \EndFor    
  \If{ BufferOut $\neq$ \{KeyArray, ValArray\} }
    \State \{KeyArray, ValArray\} $\xleftarrow{\Call{cudaMemcpy}{}}$ BufferOut
  \EndIf
\EndFunction
\end{algorithmic}
\end{algorithm}

\subsubsection{Particle Reordering}
\begin{figure}[htbp]
\centering
\includegraphics[width=6.0in]{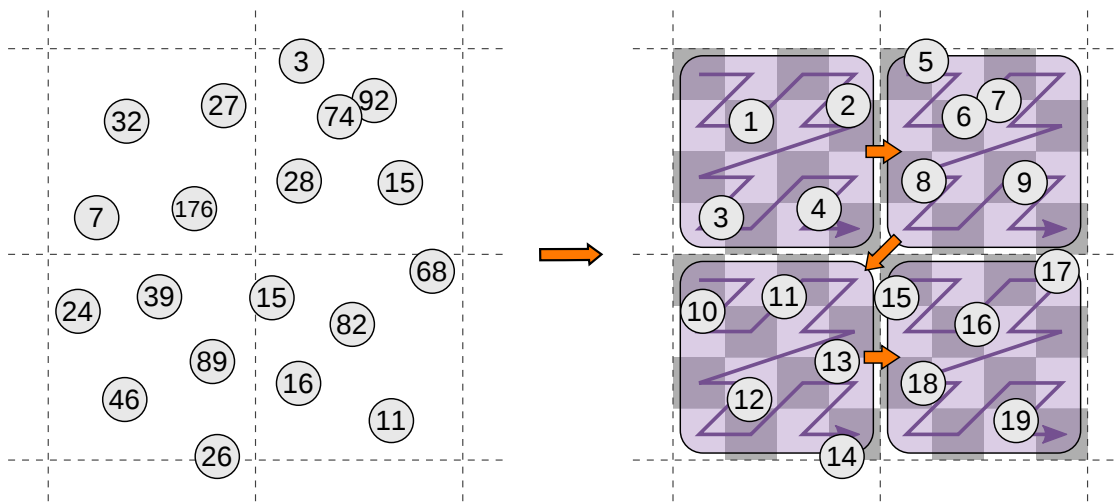}
\caption{A 2-level Morton curve is obtained by ensuring that the number of bins for reordering is a multiple of the number of bins in the cell list by some integer power of 2.}
\label{fig:reorder}
\end{figure}

\begin{figure}[htbp]
\centering
\includegraphics[width=4.0in]{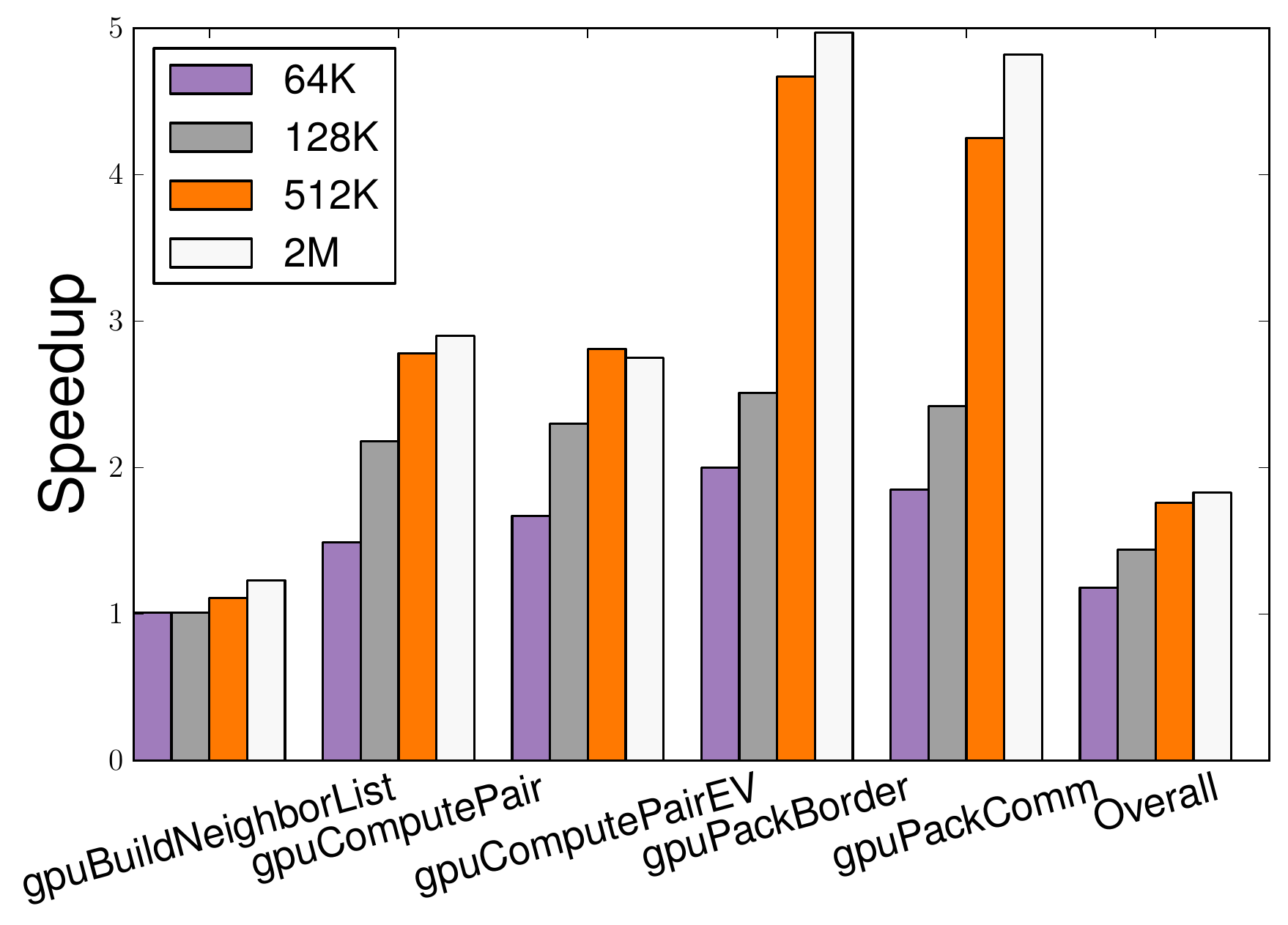}
\caption{Performance gain due to particle reordering for the whole program and individual kernels. This metric also indirectly reflects the memory bandwidth consumption for each kernel.}
\label{fig:bcmk_reorder}
\end{figure}

\begin{figure}[htbp]
  \centering
  \subfigure[reorder on]{
    \includegraphics[width=3in]{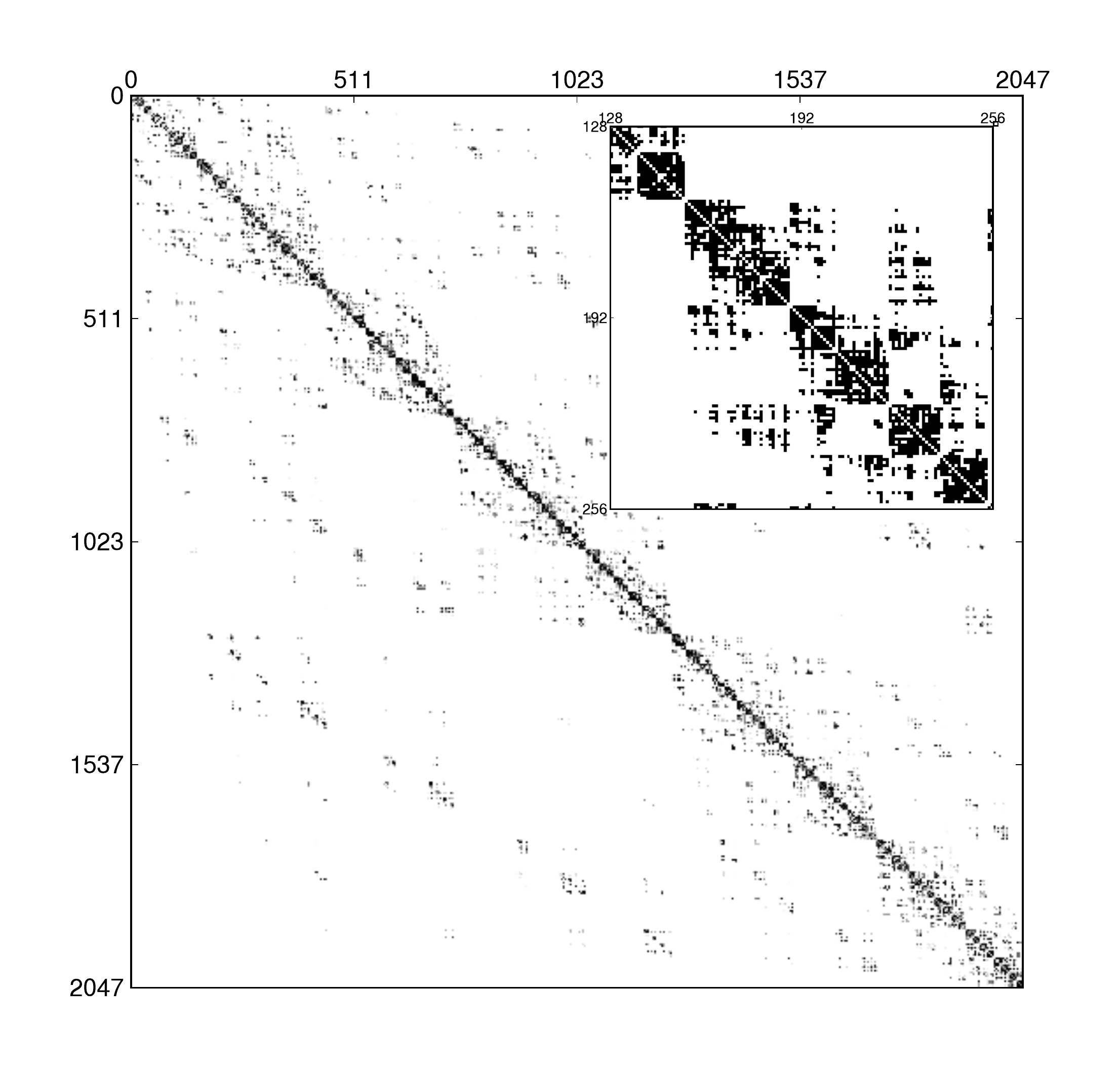}
    \label{fig:imap_y}
  }
  \subfigure[reorder off]{
    \includegraphics[width=3in]{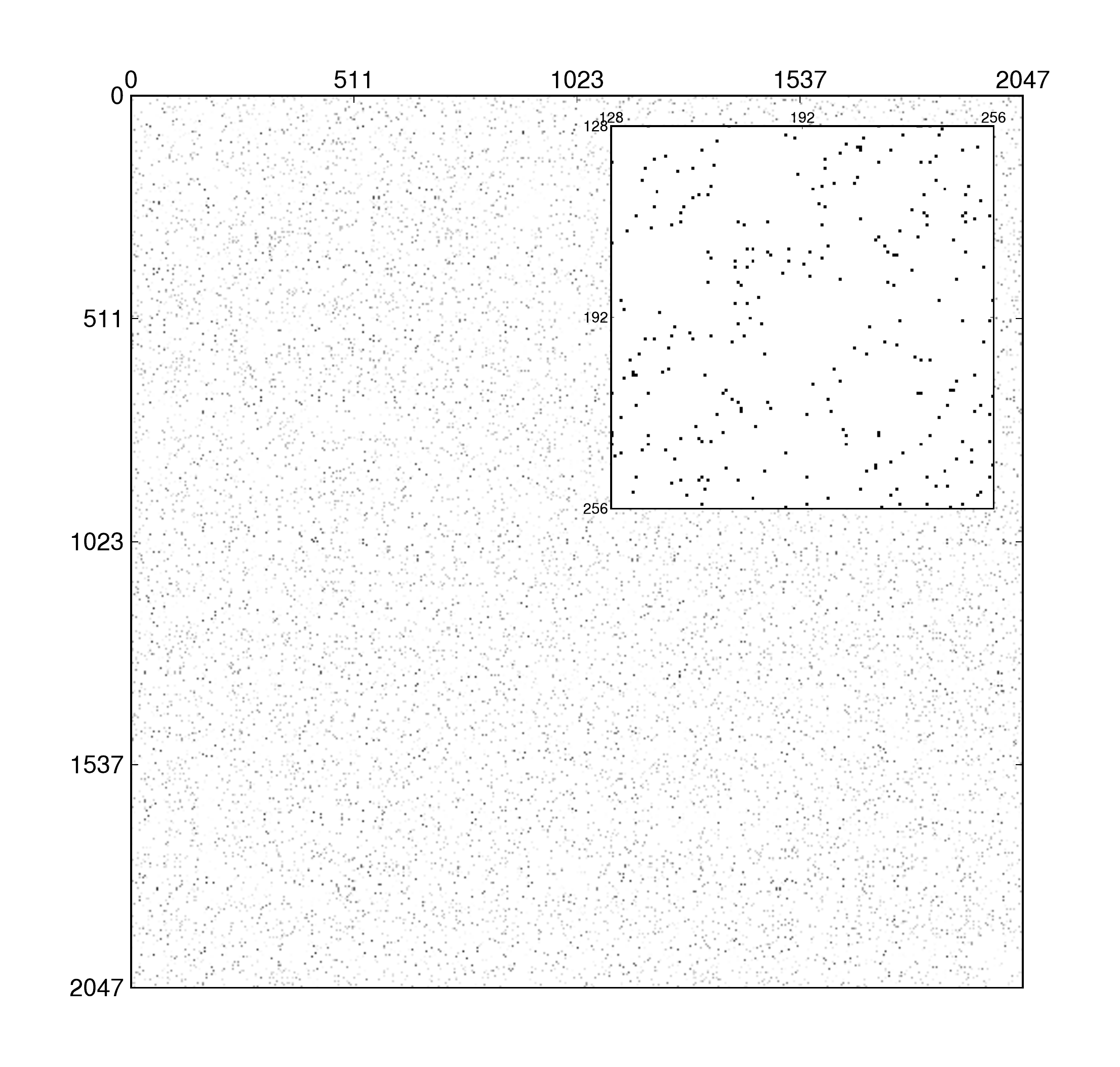}
    \label{fig:imap_n}
  }
  \caption{Interaction matrices obtained for 2048 particles with/without particle reordering. Reordering strongly diagonalizes the matrix and aggregates off-diagonal components into concentrated blocks. Particle diffusion results in randomly distributed interaction when reordering is off. }
  \label{fig:imap}
\end{figure}

Unlike grid/lattice-based methods, particle-based simulation deals with systems whose structures are largely irregular. As a result, the memory access pattern of typical particle simulation codes exhibits poor locality. The situation turns out to be worse on GPU due to the SIMD nature of the GPU architecture and a lower per-core bandwidth. In fact, most particle-based simulation programs are memory-bound, while DPD suffers even more because both coordinate and velocity data are required for evaluating the pairwise interaction. Particle reordering along a space-filling curve has been proposed as an effective technique to mitigate the problem. For example, a space-filling curve pack (SFCPACK) algorithm is used in the HOOMD-blue simulation package to sort particles on a per-block basis \cite{Anderson20085342}. Our code, on the other hand, uses a two-level Morton encoding (Z-curve) scheme to reorder particles as shown in Figure~\ref{fig:reorder}. In this scheme, cells are first sorted according to their Morton order, while at the same time particles within a cell are also sorted by their relative position along a local Morton curve in each individual cell. In addition, the boundary of the cells used for particle reordering was chosen to coincide with that used for building the cell list. The two-level Morton encoding achieves good locality while at the same time ensures particles belonging to the same cell have consecutive indices. This feature is later exploited by our neighbor list building algorithm. Note that at this point only local particles have been sorted and in order to proceed we need to obtain information of ghost particles from neighboring processors; this will be discussed in section~\ref{sec:comm}. Figure~\ref{fig:bcmk_reorder} summarizes the performance gain brought about by this particle reordering scheme for DPD fluid simulations at four different sizes compared to those without particle reordering using the same set of parameters. The overall program performance is boosted by almost 100\%, while individual GPU kernels benefit even more depending on their level of memory consumption. The effect of particle reordering is further characterized in Figure~\ref{fig:imap}, which visualizes the interaction matrices for a system of 2048 particles with and without particle reordering. While diffusion tends to randomize the distribution of interactions over the entire domain, reordering on the other hand strongly diagonalizes the matrix and aggregates off-diagonal components into concentrated blocks.

\subsubsection{Cell List}

Our program adopts the conventional approach of using the neighbor list to accelerate the evaluation of non-bond pairwise interactions. In addition, a cell list is used to further facilitate the construction of the neighbor list. The cell list construction starts with {\em binning}, \textit{i.e.} partitioning the simulation box into a number of equally sized cells and assigning particles into their corresponding cells. This can be done efficiently using parallel sorting with the particles' cell indices being the input key for sorting, while the determination of the cell index depends solely on the particle's own position and hence is embarrassingly parallel. This algorithm is employed by the GPU module of the LAMMPS package \cite{GPUDPD_LAMMPS}, which contrasts the approach in Wu's implementation where the atomic increment operation, a potential bottleneck at high parallelism, is used to build the cell list \cite{wu2011gpu}. Alternatively, since particle indices are consecutive within each cell after the reordering process, the construction of the cell list can be performed by performing particle reordering followed by detecting the cell boundary in the one-dimensional particle array without the need for further sorting of cell indices. The boundary detection is done by launching one thread per particle, checking if its cell ID differs from the particle that precedes it. In our code we adopted the latter approach.

\subsection{Neighbor List Construction}
\label{sec:neighbor}

\subsubsection{Stencil}
\label{sec:stencil}

\begin{figure}[htbp]
\centering
\includegraphics[width=2.5in]{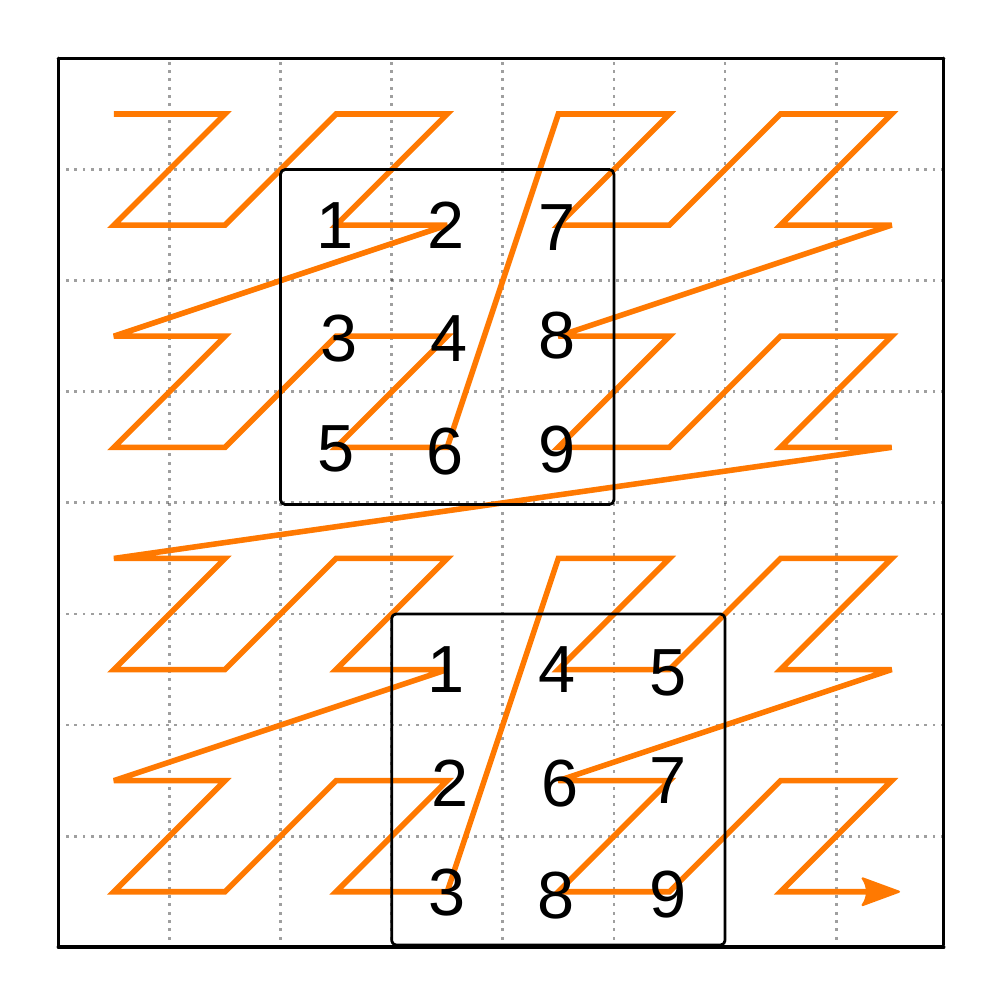}
\caption{Expanding stencils along the same Morton curve for particle reordering ensures monotonicity of the particle indices within each stencil.}
\label{fig:coar_stencil_reorder}
\end{figure}

As the first step, a coarse-grained stencil list is generated, which contains one stencil for each cell. Each stencil stores the indices of the neighboring cells surrounding its master cell. Within each stencil the cell indices are sorted according to the Morton order as illustrated in Figure~\ref{fig:coar_stencil_reorder}. Such ordering of the stencil combined with the aforementioned particle reordering technique ensures that the  particle indices in each stencil are monotonically increasing. Since the building of a coarse-grained stencil need only to be done once during the entire simulation, the sorting is simply implemented using bubble sort. The coarse-grained stencil list is then expanded into a fine-grained one, storing the actual indices of particles in each cell's neighboring cells. This is particularly useful for our DPD application because the particle density in a DPD simulation is much lower than that of an AAMD simulation. For example, a cell of volume $(r_c+\Delta r)^3$ in a DPD system contains on average 10 particles, where $\Delta r$ is the skin distance for building the neighbor list. The fine-grained stencil allows the neighboring particles from multiple cells to be loaded into our neighbor list builder in a coalesced manner without incurring branch divergence. Both the coarse-grained and the fine-grained stencils are stored in a row-major fashion.

\subsubsection{Neighbor List}

For the actual job of building the neighbor list, we invented an atomics-free algorithm that is able to generate a fully ordered neighbor list in parallel. The algorithm is deterministic, meaning that the generated neighbor list will always be the same given the same initial configuration and runtime parameter. Such deterministic approach may benefit debugging and simulation cases that require reproducibility.

The neighbor list builder works on the fine-grained stencil. Instead of the common practice of assigning one block for each cell, a warp-centric programming model, in which each cell is taken care of by a single warp, is employed to optimally balance data reuse and inter-thread communication. A fixed number of shared memory slots are assigned to each warp for storing the information of the {\em i} particles in its working cell. Each slot consists of two integers and three fp32 values for the particle index, neighbor count and the x, y, z coordinate. In practice, 32 slots are allocated to each warp to coincide with the warp size, though in theory any number can be used as long as the shared memory capacity is not exceeded. The storage needed for each slot is $ 4 + 2 \times 2 + 4 \times 3 = 20 $ bytes. On the Kepler architecture, an occupancy of 100\% translates into 2048 concurrent threads per stream multiprocessor and hence a shared memory utilization of $ 20 * 2048 / 1024 = 40 $KB, which fits well into the 48 KB on-chip shared memory. 

\begin{figure}[htbp]
\centering
\includegraphics[width=4.0in]{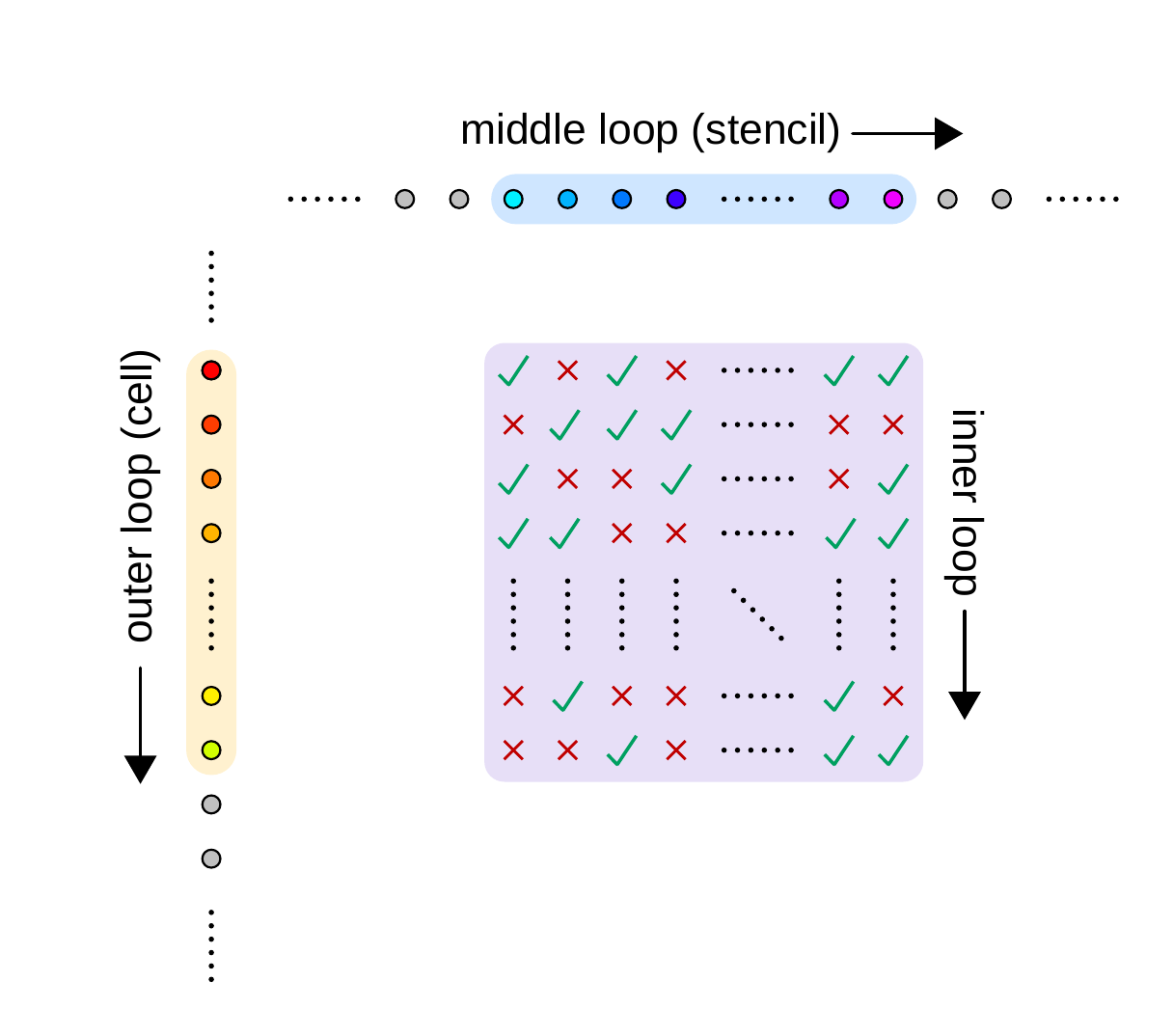}
\caption{A graphical visualization of the triple loop in the atomics-free neighbor list builder.}
\label{fig:neighbor_builder}
\end{figure}

As illustrated in Figure~\ref{fig:neighbor_builder}, each warp in our neighbor list builder works in a triple loop. The outer loop walks through the {\em i} particles in its working cell with a batch size of 32, with each thread loading one particle's index and coordinate into a slot in the shared memory. The middle loop scans over the cell's stencil with each thread loading one {\em j} particle into its private registers at a time. The inner loop performs the actual all-over-all distance checking with all threads looping through the shared {\em i} particles together. Such loop arrangement minimizes the {\em long tail} effect because branch divergence occurs only in the last iteration of the middle loop if the stencil length is not a multiple of the batch size. Typical stencils contain 200 particles each so the performance hit due to a branched last iteration is not significant.

Because multiple {\em j} particles may be found staying within the cutoff distance of one {\em i} particle at the same time, a commit of the {\em j} index into the correct position of the neighbor list intuitively requires the use of atomic increment operation to resolve potential conflict. This can also be thought of as each thread needs the global hit/miss information from other threads to determine the insertion point for its own commit. However, realizing that in our algorithm at any given time there is only one warp evaluating the neighbors for a given particle, we can reduce the hard problem of global information gathering into a simpler one of collecting a boolean value from a warp. The warp vote function, \_\_ballot(), fits naturally for this job by enabling fast communication between the threads and eliminating the use of atomic instructions. The ballot intrinsics takes in a predicate value from each active threads of a warp and returns to each thread an unsigned integer whose N-th bit is set if the predicate evaluates to true for the N-th thread. In the inner loop of our neighbor list builder, each thread ballots its distance checking predicate to form a bit vector \texttt{nHit}. Shifting \texttt{nHit} by \texttt{warpSize} - \texttt{laneId} toward left, where $\texttt{lane id} \in [0..\texttt{warpSize}-1]$ is the ordinal number of a thread within a warp, and then counting the number of remaining set bits in the bit vector tells a thread the number of hits made by threads with a lower \texttt{lane id} than itself. The insertion point is then simply determined by adding this number to the existing number of neighbors for the {\em i} particle from previous iterations. The number of hits from an iteration is then accumulated to the total neighbor count for the {\em i} particle by a single thread, in our case lane 0, after the cooperative neighbor list insertion. In contrast to the atomic-based approach, this warp vote-based approach for committing particle indices into the neighbor list is completely deterministic and only involves fast intra-warp communication. Recall that the indices in the stencil are monotonically increasing, this implies that the neighbor list generated in this way are also strictly increasing for each particle, yet no sorting was performed explicitly on the neighbor list. A careful benchmark shows that our atomics-free builder executes twice as fast on average compared to an equivalent  builder using atomic increment as shown in Table~\ref{table:neigh_builder}.

\begin{algorithm}
\caption{Atomics-free neighbor list builder}
\label{ALG_Neighbor_builder}
\begin{algorithmic}[1]
\Function{gpuBuildNeighborList}{}
  \For{each cell \textbf{in parallel}}
    \State \textbf{shared} slot[warpSize]
    \State l $\gets$ threadIdx \textbf{mod} warpSize
    \State n $\gets$ number of particles in the cell
    \State p $\gets$ 0
    \LComment{load working cell}
    \While{ p $<$ n }
      \State part $\gets$ \Call{Min}{ n - p, warpSize }
      \If{ l $<$ part }
        \State slot[l].i $\gets$ cell[ p + l ]
        \State slot[l].r $\gets$ \Call{Texture}{coordinate, slot[l].i }
        \State slot[l].nn $\gets$ 0
      \EndIf
      \LComment{load stencil}
      \For{ k $=$ l \textbf{to} nStencil \textbf{by} warpSize \textbf{in parallel}}
        \State j $\gets$ stencil[k]
        \State r $\gets$ \Call{Texture}{coordinate,j}
        \LComment{distance check}
        \For{ i $=$ 0 \textbf{to} part \textbf{by} 1} 
          \State dr$^2$ $\gets$ ( r - slot[i].r )$^2$
          \State hit $\gets$ dr$^2$ $<$ r$_c^2$ ~?~ \textbf{true} : \textbf{false}
          \State nHit $\gets$ \Call{Ballot}{hit}
          \State nAhead $\gets$ \Call{ShiftLeft}{nHit, warpSize - l}
          \State pIns $\gets$ slot[i].nn + \Call{Popc}{ nAhead }
          \If{ hit }
            \State neighborList[ slot[l].i ][ pIns ] $\gets$ j
          \EndIf
          \State accumulate \Call{Popc}{nHit} to slot[i].nn by lane 0
        \EndFor
      \EndFor
      \If{ l $<$ part }
        \State store slot[l].nn to global storage
      \EndIf
      \State p $\gets$ p + warpSize
    \EndWhile
  \EndFor
\EndFunction
\end{algorithmic}
\end{algorithm}

\begin{figure}[htbp]
\centering
\includegraphics[width=4.0in]{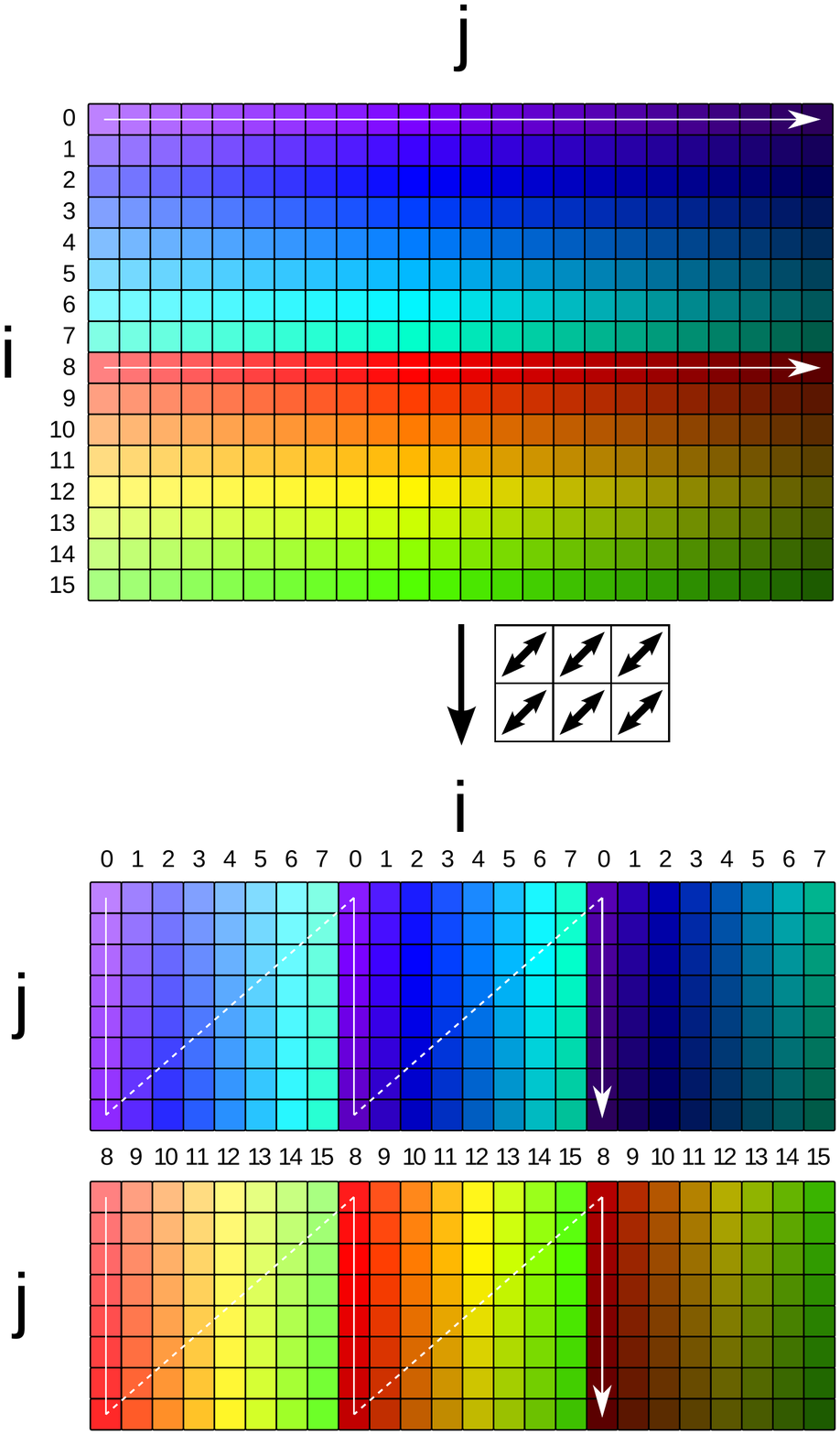}
\caption{Local in-place transposition of the neighbor table. The memory footprint for a single thread when looping through its neighbors as indicated by the arrowed lines produces fully coalesced access.}
\label{fig:local_trans}
\end{figure}

Another issue to be considered here is the layout of the neighbor table. A row-major layout assigns one consecutive line of memory for each particle, while a column-major layout, on the other hand, stores the n-th neighbors of all the particles consecutively in the n-th row. On the GPU we typically assign each thread a particle when evaluating the pairwise interaction, which implies that a column-major layout is more efficient because fetching such a list results in fully coalesced memory loads. However, a column-major table is much less efficient when being written to because an entire warp works on the same {\em i} particle in the inner loop in our neighbor list builder. In this case, each insertion of the {\em j} indices from a thread generates a write request of 32 bytes, among which only 4 bytes are actually useful data, resulting in a memory efficiency merely 12.5\%. To combine the advantages from both layouts, we adopt a write-transpose-read approach in which the neighbor table is built in a row-major fashion, and then transposed for maximum loading efficiency in the subsequent computation. In practice, we only {\em locally in-place} transpose each 32 by 32 tile of the neighbor table matrix. We made this compromise because a full transposition of such a large matrix is complicated with suboptimal efficiency \cite{libmarshal}. A schematic representation of the local transposition is presented in Figure~\ref{fig:local_trans}.

Further, a double-insertion trick can be utilized by the neighbor list builder to help suppress branch divergence in the force kernel. This heuristic depends on the assumption that particles which were within the actual cutoff distance when building the neighbor list are more likely to be found within the cutoff distance again in subsequent time steps than those which initially lie between the cutoff and the skin distance. To separate the two types of neighbors in the neighbor list, the warps in our neighbor list builder can be modified to perform two commits per inner loop iteration, one for the {\em core} particles, and the other one for the {\em skin} particles. The {\em core} particle indices are stored in the original order, while the {\em skin} particle indices were stored at the back of each row in reversed order. The neighbor count integer used in the shared slots in the neighbor list builder is also split into two short integers correspondingly. In this way we still need only one table for storing the neighbor list. A separate kernel is used to join the two parts together, making the splitting trick completely transparent to the rest of the code. Note that the naive way of simply forcing every thread to enter the force evaluation branch regardless of the actual distance results in worse performance, since it issues more stress on the memory bandwidth and the texture units. It is even erroneous in DPD since the functional form does not accept $|\textbf{r}_{ij}|$ greater than $r_c$.

\subsection{Precision Model}
\label{sec:prec}

We use a hybrid precision model for the neighbor list builder and force evaluation kernel to optimize GPU memory bandwidth usage. Force evaluation in DPD is expensive in terms of both memory bandwidth consumption and arithmetic complexity. The memory bandwidth consumption is high because both the coordinate and the velocity of the interacting particles are involved in the functional envelope. For each pair of interaction an input of 2 integer and 6 fp64 values, or a total of 56 bytes, is required to supply the required information. We used the standard approach of texture data mapping to alleviate the problem. Note that coordinate and velocity are stored in double precision in our code everywhere else, but they are converted to single precision before being mapped as textures for force evaluation. Specifically, the single-precision coordinates are casted from the double-precision difference between the actual particle coordinates and the geometric center of the owning processor's compute domain. The double-precision velocity is cast to single precision directly. Our approach differs from that of G{\"o}tz \textit{et al} \cite{Amber_SPFP} in that the computation is still done in full double precision. The neighbor list builder works entirely in single precision because the distance comparison is not sensitive to precision. The possibility of using the newly-introduced non-coherent cache pathway to load such data was also explored. However, a microbenchmark revealed that on the current Kepler GPU the latency for a cache hit in the non-coherent cache (140 cycles) is much longer than that of a texture fetch (108 cycles), though is still shorter than that of a regular global memory access (220 cycles) \cite{microbenchmark}.

\subsection{Numerical Optimization}
\label{sec:numerical}

The arithmetic complexity of the DPD pairwise interaction is high for two reasons. First, for each pair of the random force $F_{ij}^R$, one Gaussian random number has to be consumed. This translates into one square root, one natural logarithm, one trigonometric function and the generation of two uniform random numbers assuming the use of the Box-Muller algorithm \cite{Box-Muller}. Second, the power function is needed to evaluation the weight function $w_R(\textbf{r}_{ij}) = w_C^s(\textbf{r}_{ij}) = (1-r/r_c)^s$ if the exponent $s$ takes a value other than positive integers.

Using of a separate RNG library to generate the random numbers beforehand and storing them in double precision requires a storage space that is twice the size of the neighbor list. Besides, it is inevitable for pairs of interacting particles to get separated onto different processors during domain partitioning. In this case, we either have to compute the pairwise interaction only once on one of the processors and then communicate the force back, or we have to reproduce the same random number on different processors to save the communication. Generally, communication is expensive for the GPU so it is preferred if the random numbers can be reproduced. This dictates that we use lightweight hash function-based generators and particle properties that are invariant across different nodes as the seed to generate the random numbers \textit{in situ}. In particular, the tiny encryption algorithm (TEA) as shown in Algorithm~\ref{ALG_TEA} has been proposed as a suitable choice for DPD simulation with the input seeds chosen to be a mixture of the particle indices, current time step and a global seed \cite{TEA,zafar2010gpuTEA,TEA_DPD}.

\begin{figure}[htbp]
\centering
\includegraphics[width=4.0in]{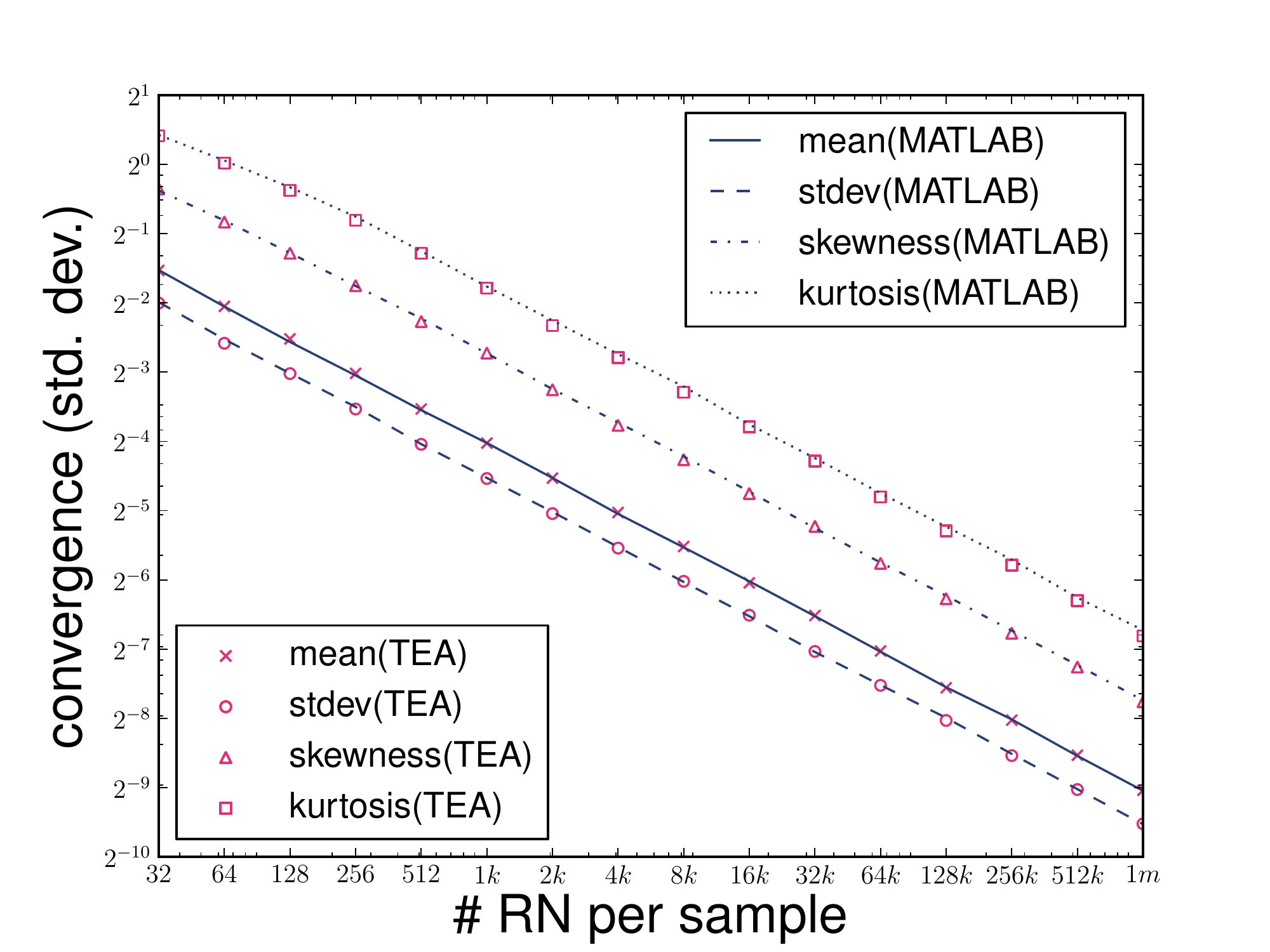}
\caption{The convergence of the distribution of the random numbers generated by our signature-based TEA is compared to that generated by MATLAB using up to the 4-th moment.}
\label{fig:tea}
\end{figure}

The quality of the random numbers generated by TEA increases with the number of iterative rounds performed. Due to the limited sampling range of the input seeds in DPD (most simulations contain no more than a few million particles), at least 8 rounds of hashing are required to obtain sufficient randomness. We took a different approach of sampling from a wider source of entropy for the input seeds to guarantee randomness rather than simply increasing the number of iterative rounds. In particular, for each particle we take the bit-reversed particle identity tag as the first integer, and an interleaving of the first 11 bits of the mantissa of the three-dimensional velocity components as the second integer. Besides, we employ a pre-processing step to blend the two integers using 16 rounds of the TEA hashing. The two integer output from this preprocessing step are then combined using bitwise exclusive or into a single 32-bit binary signature for every particle. The overhead of such preprocessing is negligible since it is embarrassingly parallel with complexity $\mathcal{O}(N)$. Only 4 rounds of TEA hashing are needed when using the preprocessed binary signature. In order to validate this approach, we dumped the random numbers generated during one simulation of 65,536 DPD particles and compared up to the 4-th moment of the numbers against that generated by MATLAB. The result matches perfectly as shown in Figure~\ref{fig:tea}.

\begin{algorithm}
\caption{The Tiny Encryption Algorithm \cite{TEA}} and our signature generator.
\label{ALG_TEA}
\begin{algorithmic}[1]
\Function{TEA}{$n,v_0,v_1$}
\State $ sum \gets 0 $
\For{ $n$ rounds }
  \State $ sum \gets sum + delta$
  \State $ v_0  \gets v_0 + \Call{BitXor}{ \Call{ShiftLeft}{v_1,4} + k0, \Call{ShiftRight}{v_1,5}+k1, v_1 + sum } $
  \State $ v_1  \gets v_1 + \Call{BitXor}{ \Call{ShiftLeft}{v_0,4} + k2, \Call{ShiftRight}{v_0,5}+k3, v_0 + sum } $
\EndFor
\EndFunction
\\
\Function{Preprocess}{$i$}
\State $ v_0 \gets \Call{BitReverse}{\mathrm{Tag}[i]} $
\State $ b_x \gets \Call{GetMantissa}{\mathrm{Velocity}[i].x,11} $
\State $ b_y \gets \Call{GetMantissa}{\mathrm{Velocity}[i].y,11} $
\State $ b_z \gets \Call{GetMantissa}{\mathrm{Velocity}[i].z,11} $
\State $ v_1 \gets \Call{Interleave}{b_x,b_y,b_z} $
\State $ \Call{TEA}{16,v_0,v_1} $
\State \Return $ \Call{BitXor}{v_0,v_1} $
\EndFunction
\end{algorithmic}
\end{algorithm}

The uniform random numbers generated by the TEA algorithm above are converted to Gaussian using the Box-Muller equation. Profiling indicated that a plain implementation of the Box-Muller conversion alone consumes about 25\% of the total time for pairwise force evaluation. This is actually not surprising since it involves the computation of transcendental functions as mentioned previously. However, there are facts that we can exploit to accelerate the math functions. First of all, we have perfect control over the range of input for the functions, \textit{i.e.} the uniform random numbers are always between 0 and 1. In addition, even though we are producing double precision floating points as the final output, it does not necessarily mean that we need to compute in the full precision during the intermediate steps. By looking at the Box-Muller equation $ z_1 = \sqrt{-2 \ln{u_1}} \cos(2\pi u_2) $, it is clear that the radial component of the number depends only on $u_1$, while the phase component depends only on $u_2$, both of which contain only 32 bits of entropy as specifies by the TEA algorithm. It is through the combination of the two parts that we obtain a result of full double precision.

Hence, we implement our own reduced-precision natural logarithm and cosine as follows:

a) Instead of converting the 32-bit unsigned integer $v_1$ into a double precision floating point number $u_1 \in [0,1)$ and then performing the natural logarithm, we apply a binary logarithm directly to $v_1$. The evaluation can be split into the integral part and the fractional part as $ \log_2{v_1} = \log_2{ 2 ^ {I+F} } = I + \log_2{ 2^F }$, where $I$ is an integer and $F$ is a fractional number between 0 and 1. The integer part $I$ can be obtained trivially using the relation $ I = 31 - clz(v_1) $, where $clz$ stands for {\em consecutive leading zeros}, which is a hardware-implemented PTX instruction on the Kepler GPU. The fractional part is approximated with an 8-th order Chebyshev polynomial as $ \log_2{x} = z \times P(z) $, where $ z = \frac{x-1}{x+1}$ \cite{comp_approx_bible}. The natural logarithm is derived by subtracting the binary logarithmic result by 32 and then multiplying by $\ln{2}$. Our implementation compiles into 47 branch-free PTX instructions, which contrasts the CUDA native implementation that compiles into 89 PTX instructions with 5 conditional branches and one floating point reciprocal. The maximum relative error for our implementation is $4.21\times 10^{-12}$ over the entire range of unsigned integer, yielding a binary precision of 42.8 bits.

b) In the custom cosine function, the highest bit of $v_2$ is taken as a sign bit, while the low 31 bits are converted into a double precision floating point number $ u_2 = v_2[0..30] \times 2^{-31} \in [0,1)$. This simplifies the range reduction for the evaluation of cosine, enabling us to approximate the function with a 11-th order Chebyshev polynomial, which has only 6 terms thanks to the axial symmetry of the cosine function. The resulting function compiles into 23 branch-free PTX instructions using only cheap arithmetics such as bit shifting, floating point multiply and fused multiply-addition. The CUDA native implementation, on the other hand, compiles into 54 PTX instructions with 3 conditional branches and potential conflict in the constant cache. Our implementation gives a maximum relative error of $1.10\times 10^{-10}$ over the range [0,1], or a binary accuracy of 44.1 bits.

The double-precision power function used in evaluating the weight function $w_R(\textbf{r}_{ij})$ is one of the most time-consuming functions in the CUDA math library. In order to conform to the IEEE floating point standard, the CUDA native implementation has to deal with the full range of inputs as well as possible exceptions. As a result, the native power function compiles into 258 PTX instructions with 18 conditional branches, 9 uniform branches, 3 reciprocal and 1 division with an average execution latency of 1982 clock cycles. Again in our implementation we seek to exploit the knowledge over the input range for optimization: the exception conditions that the base or the exponent being 0 can be precluded by the cutoff testing prior to the function call; it is also unlikely that the base or the exponent would be NaN or Inf unless there are serious problems in the underlying physics of the model. Our custom double precision power function is based on the base-2 logarithm and exponential using the identity $ a^b = 2 ^ { b \mathrm{log_2}a } $ as shown in Algorithm~\ref{ALG_POW}. The logarithm part is similar to that used in our Box-Muller implementation except that the order of the Chebyshev polynomial is increased to 14. The exponential part is approximated by a 11th order Chebyshev polynomial. Both routines give an maximum error of less than 1 unit in the last place (ULP). Special care has been taken to ensure the accuracy of the composite function when combining the results. The maximum error of our power function is 6 ULP given an input range of $ a \in [10^{-102},10^{102}] $ and $ b \in [0,3] $, or 11 ULP given an input range of $ a \in [10^{-51},10^{51}] $ and $ b \in [0,6] $. This should suffice for the purpose of DPD simulation, where typically $a \in [10^{-10},2]$ and $ b \in [0.25,3.0] $.

\begin{algorithm}
\caption{Branchless power function with error $\leq$ 11 ULP. The \texttt{\_\_hiloint2double}, \texttt{\_\_double2hiloint} and \texttt{\_\_fma} functions are double-precision floating point intrinsics from the CUDA math library.}
\label{ALG_POW}
\begin{lstlisting}
// fast computing for integer power of 2
double power2( int n ) { return __hiloint2double( (1023+n)<<20, 0 ); }

// 11th order 2^x = P(x)
// error < 1 ULP for x in [0,1.0]
double exp2_frac( double x ) {...}

// fast converging 14-order log2(x) = z*P(z^2)
// error < 1ULP for x in [1,2]
double log2_frac( double x ) {...}

double fastpow(double a, double b)
{
  int hi, lo;
  __double2hiloint( a, hi, lo );
  // extract exponent
  double I  = ( hi >> 20 ) - 1023;
  // reset exponent, do fractional log
  hi = ( hi & 0X000FFFFF ) | 0X3FF00000;
  double F  = log2_frac( __hiloint2double(hi,lo) );
  // multiply by exponent, separate to 2 parts to mantissa precision loss
  double II = floor( b * ( I + F ) );
  return power2( II ) * exp2_frac( __fma( b, F, __fma( b, I, -II ) ) );
}
\end{lstlisting}
\end{algorithm}

\subsection{Communication}
\label{sec:comm}

In our code, each processor maintains a {\em sendlist} for every neighboring processors recording the indices of local particles whose information is to be sent during the communication. The determination of the border particles can be implemented relatively straightforward on CPUs, whereas on GPUs it requires the cooperation of multiple kernels. Nevertheless, we still implemented it on GPU because the border determination must be done after particle reordering where particle indices are reassigned. A total of seven kernels are used as outlined in Algorithm~\ref{ALG_BORDER} for border determination. Communication of ghost particle position and velocity between two border-determination steps involves only the last three steps of the algorithm.

\begin{algorithm}
\caption{Exchange of border information.}
\label{ALG_BORDER}
\begin{algorithmic}[1]
\Function{BorderDetermination}{}
  \State \textbf{flag} particles to be sent using a 1-bit boolean                         
  \State multi-block segmented prefix sum: \textbf{down-sweep} phase                      
  \State multi-block segmented prefix sum: \textbf{blocks}                                
  \State multi-block segmented prefix sum: \textbf{up-sweep} phase                        
  \State send list generation: \textbf{parallel compaction} using scan result             
  \State \textbf{pack} information of particles in send list                              
  \State MPI communication...
  \State \textbf{unpack} list of incoming ghost particles                                 
\EndFunction
\end{algorithmic}
\end{algorithm}

\section{Code Verification \& Benchmark}
\label{sec:benchmark}

\subsection{Method}

Benchmarks were carried out on the TITAN supercomputer located at the Oak Ridge National Laboratory \cite{titan2012}. The TITAN is a Cray XK-7 system with 18688 nodes. Each node is equipped with one AMD Opteron 6274 CPU and one NVIDIA Kepler K20X GPU. Each Opteron 6274 CPU has 16 integer cores and 8 floating point cores running at 2.2 GHz along with 16 MB of L3 cache. The Kepler K20X, with 2688 CUDA cores, has a theoretical double-precision performance of 1.31 TFLOPS and a peak memory bandwidth of 250 GB/s. Only the main loop is timed in the benchmark since initialization and clean up overheads are neglectable for long-time simulations. The CPU code for reference is a vanilla version of LAMMPS with our own DPD pair style implementing the aforementioned functional form. The polar form of the Box-Muller method is used for generating Gaussian random numbers on the CPU because it has much better performance than the basic form when used on the CPU. All CPU code is compiled with the Intel compiler using -O3 optimization, while GPU code is compiled by the NVIDIA NVCC compiler.

\subsection{Flow Simulation}

To verify the correctness of our code, we aimed at reproducing the fluid viscosity measured from the double Poiseuille flow used by Backer \textit{et al} \cite{backer2005}. The parameters are chosen as $ \sigma = 4.5 $, $ \rho = 6.0 $, $ k_B T = 0.5 $ and $ \delta t = 0.001 $. The conservative force is left out in the same way as in the original paper. A total of 4608 particles were simulated in a simulation box of dimension $ 12 r_c \times 8 r_c \times 8 r_c $ in the x, y and z directions, respectively. A body force $ g_z = 0.055 $ was applied to drive the flow in the periodical space.  The viscosity of the fluid was evaluated from 10 parallel simulations with different initial configurations and random seeds. We obtained a viscosity of $ 2.089 \pm 0.009 $, which is in excellent agreement with the published value of $ 2.09 \pm 0.02 $.

\begin{figure}[htbp]
\centering
\includegraphics[width=4.0in]{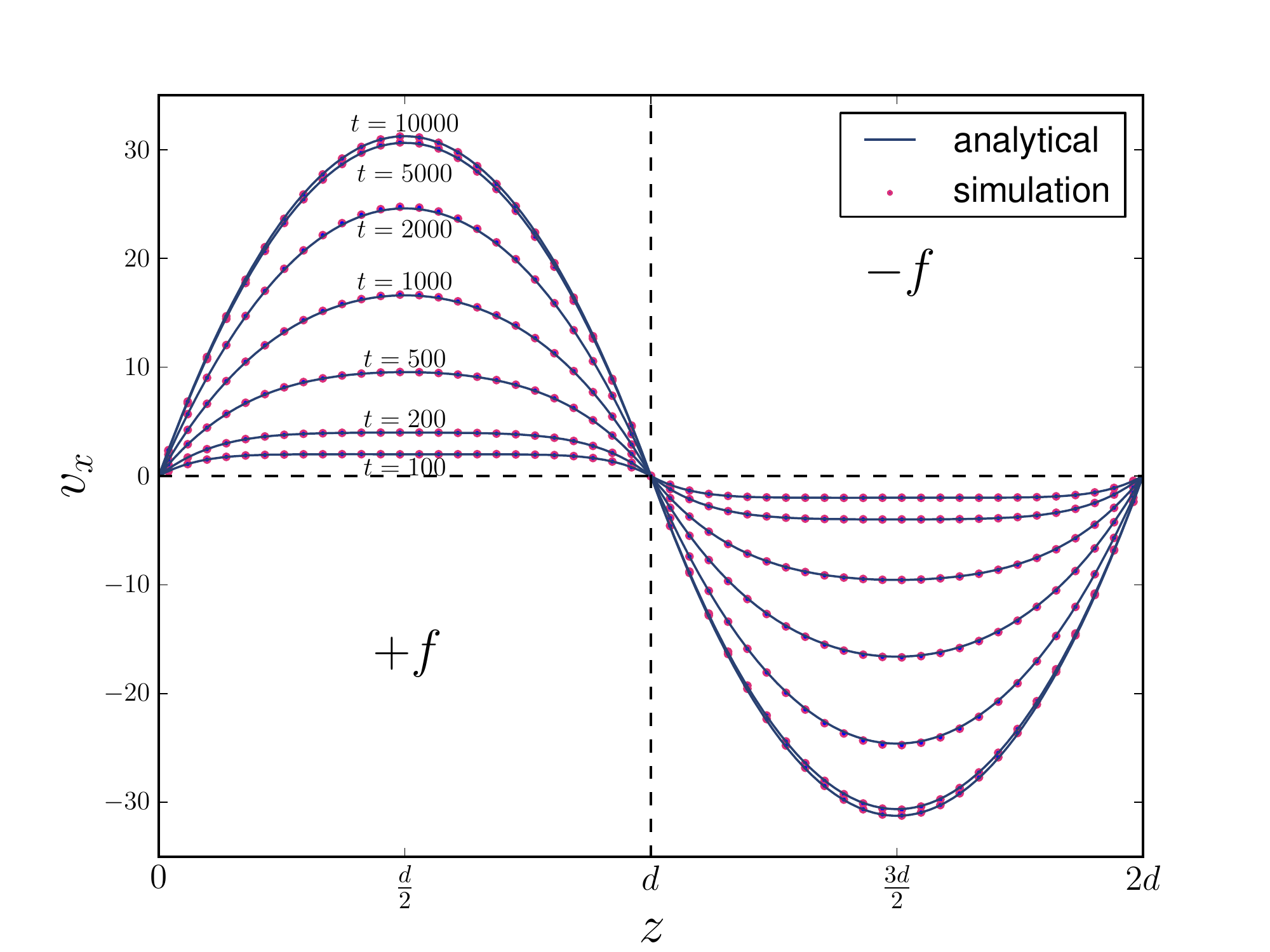}
\caption{Time evolution of the velocity profile for the double Poiseuille flow.}
\label{fig:velocity_profile}
\end{figure}

Another test case of {\em transient} double Poiseuille flow was employed to examine the accuracy of our code over long-time integration. The parameters are chosen as $ \rho = 5.0 $, $ a_{ij} = 15.0 $, $ \sigma = 3.0 $, $ k_B T = 1.0 $, $ r_c = 1.0 $ and $ \delta t = 0.01 $. The system consists of 262,144 particles in a simulation box of $ 59.4123 \times 7.42654 \times 118.825 $ in the x, y and z directions. Velocity profiles along the z direction were collected for time T = 100, 200, 500, 1000, 2000 and 10000. For each time point the velocity profile is sampled from a time window of $ [T-0.5,T+0.5] $. The simulation was repeated 10 times with different initial configurations and random seeds. The simulation gives good matching with the analytic solution given in Eq.~\ref{EQN_V_t} by Sigalotti \textit{et al.} \cite{sigalotti2003}.
\begin{equation}\label{EQN_V_t}
u(z,t) = \frac{{F{d^2}}}{{8\upsilon }}\left( {1 - {{\left( {\frac{{2z}}{d}} \right)}^2}} \right) - \sum\limits_{n = 0}^\infty  {\frac{{4{{\left( { - 1} \right)}^n}F{d^2}}}{{\upsilon {\pi ^3}{{\left( {2n + 1} \right)}^3}}}}  \cdot \cos \left[ {\frac{{\left( {2n + 1} \right)\pi z}}{d}} \right] \cdot \exp \left[ { - \frac{{{{\left( {2n + 1} \right)}^2}{\pi ^2}\upsilon t}}{{{d^2}}}} \right]
\end{equation}
A comparison of analytical and DPD simulation results is shown in Figure~\ref{fig:velocity_profile}.

\subsection{Single Node Speedup}

\begin{figure}[htbp]
\centering
\includegraphics[width=4.0in]{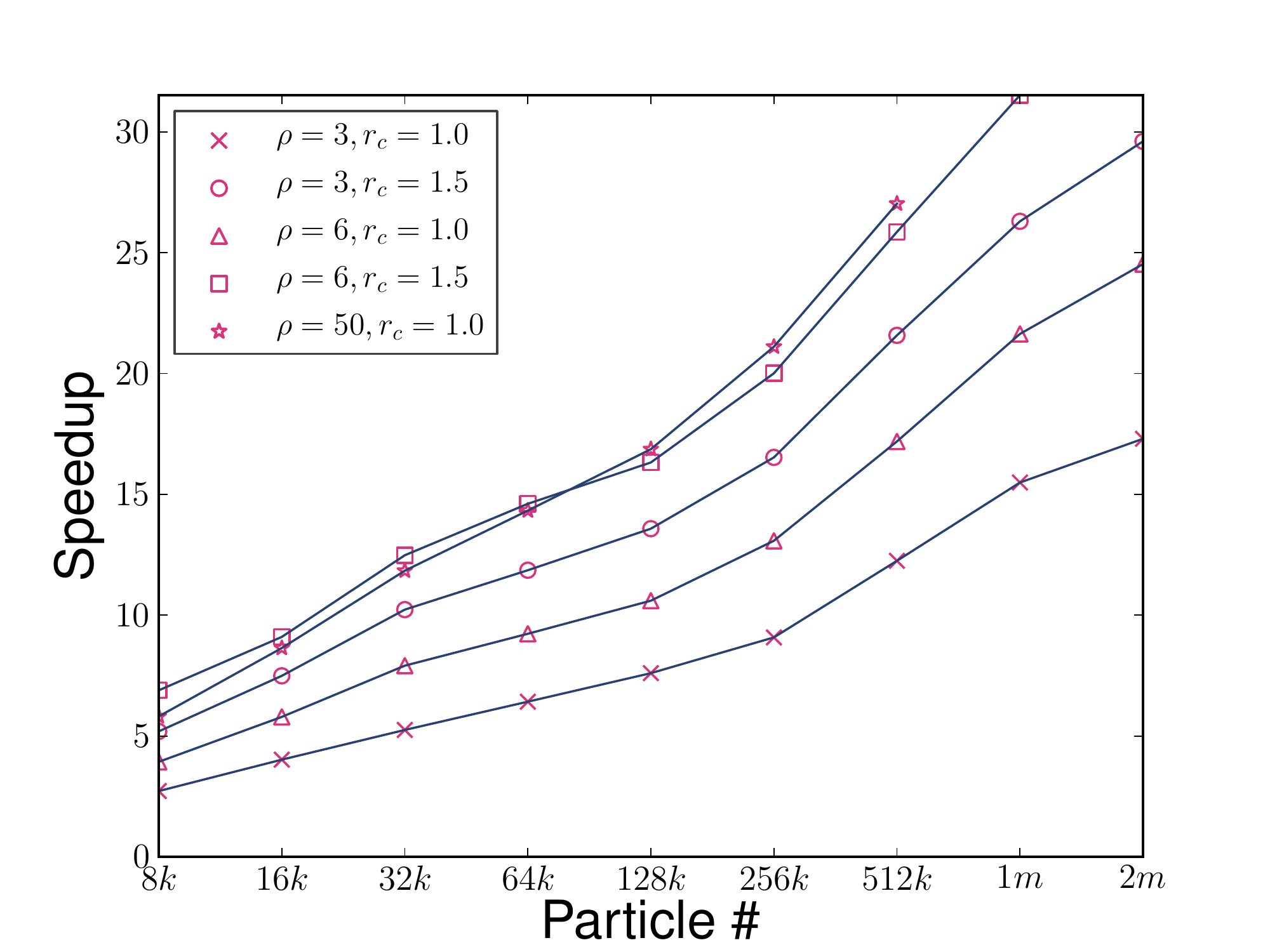}
\caption{Speedup of our code on a single Kepler K20x GPU over 16 AMD Opteron 6274 CPU cores for various system size, cutoff distance and particle number density.}
\label{fig:scaling_intra}
\end{figure}

Three different particle densities, 3, 6, and 50, and two cutoff distances, 1.0 and 1.5, were used to fully characterize the performance of our code. Note that a particle density of 50 is only observed for atomistic systems, yet such comparison allows us to estimate the performance of our algorithm when applied to an atomistic system. A time step of 0.005 was used, while the neighbor list was updated every 5 time steps. Systems whose size ranged from 8192 to 2 million particles were simulated. To enable a fair competition between the GPU and CPU implementations, the speedup is measured for one Kepler GPU over all available CPU cores on a single node. Speedup is measured as the ratio of wall time elapsed for carrying out a specific simulation. As shown in Figure~\ref{fig:scaling_intra}, the GPU code generally performs better for larger systems, with the best speedup of over 30 measured with 2 million particles. This is not surprising since overhead, especially kernel launching latency, is independent of the system size. Another interesting fact is that the speedup increases rapidly once the system size grows over 200,000. We speculate that the CPU may be experiencing cache depletion above this point.

\subsection{Weak and Strong Scaling}

\begin{figure}[htbp]
\centering
\includegraphics[width=4.0in]{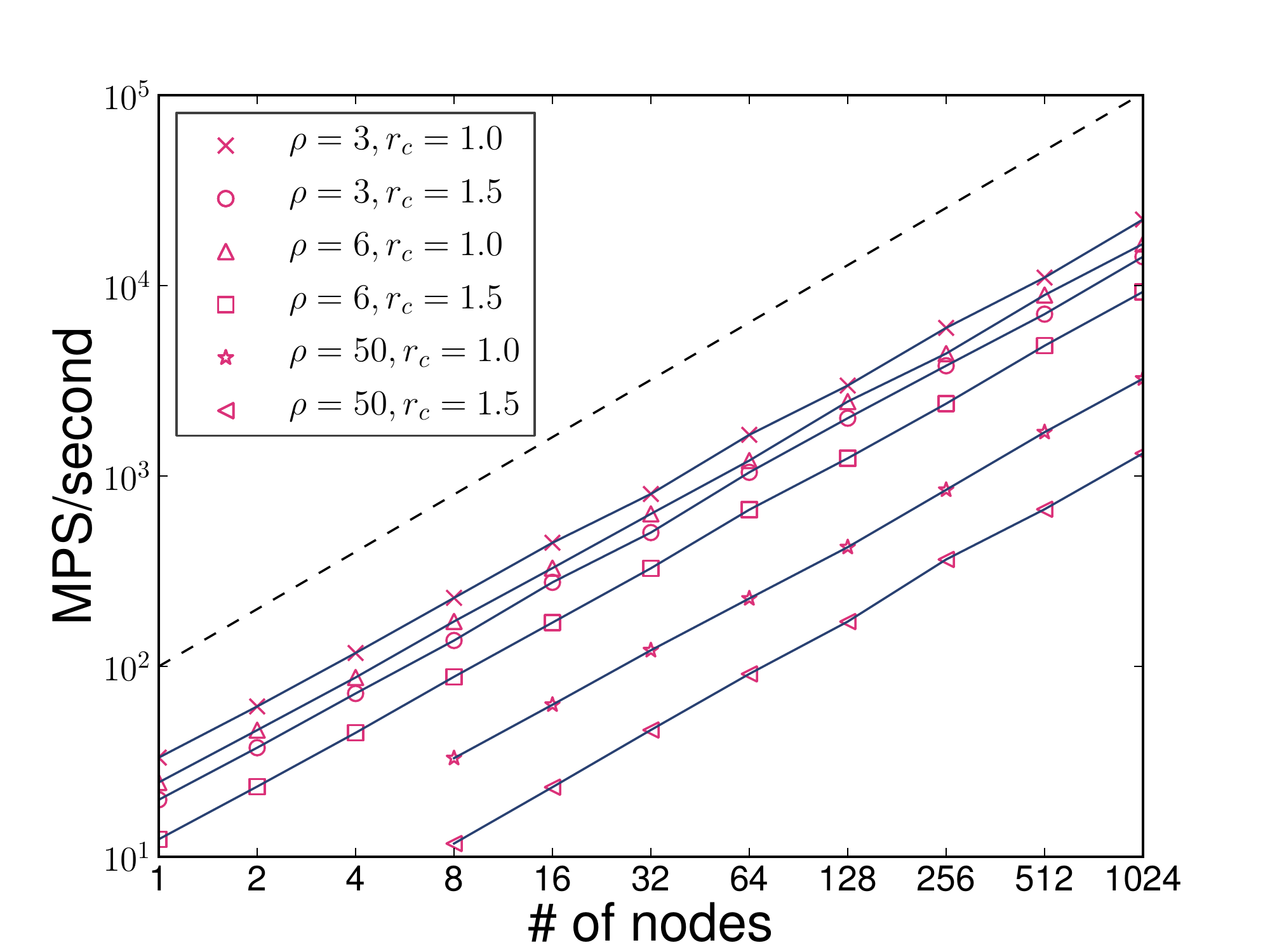}
\caption{Weak scaling performance of our code for various cutoff distance and particle number density. The system size is kept at 1 million particles per node.}
\label{fig:scaling_weak}
\end{figure}

The same set of test cases from the single node benchmark were used to establish the weak scaling and strong scaling benchmarks. A more transferrable metric, million particles$\cdot$steps per second, or {\em MPS/second}, is used to characterize the performance. For weak scaling, particles per node was kept at 1 million for $\rho = 3$ and $5$, and was kept at 128k for $\rho=50$. A nearly linear speedup is observed for up to 1024 nodes as demonstrated by Figure~\ref{fig:scaling_weak}.

\begin{figure}[htbp]
\centering
\includegraphics[width=4.0in]{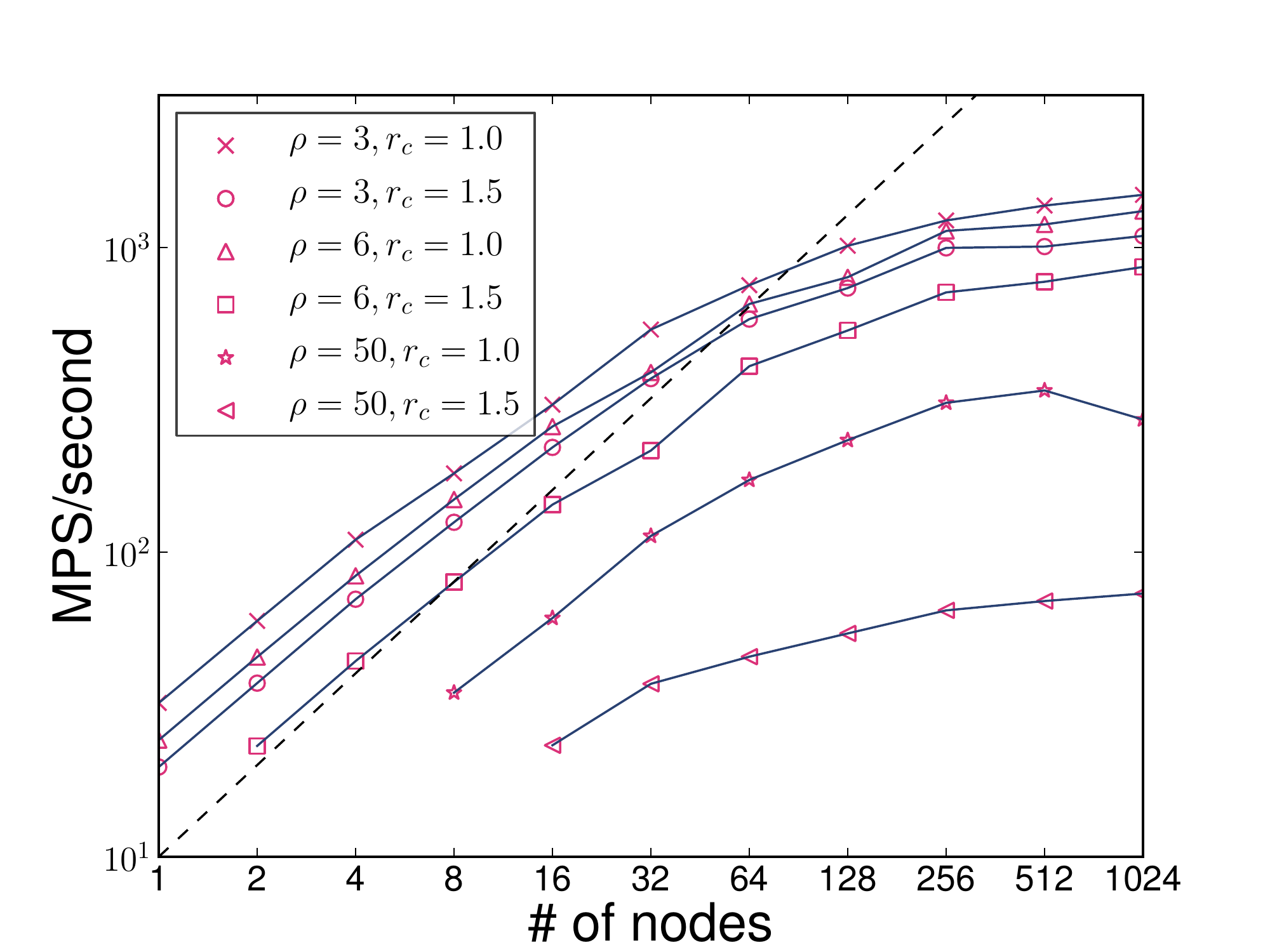}
\caption{Strong scaling performance of our code for various cutoff distance and particle number density. The system size is fixed at 2 million particles regardless of the node number.}
\label{fig:scaling_strong}
\end{figure}

Systems containing 2 million particles at different particle densities were simulated for the strong scaling benchmark. The result is shown in Figure~\ref{fig:scaling_strong}. Parallel efficiency is satisfactory for up to 64 nodes, but then it deteriorates gradually. The efficiency declination is attributed primarily to the overhead of copying data between the main memory and the GPUs when preparing the MPI communication. An attempt to optimize such latency using the newly-introduced GPUDirect RDMA pathway combined with the CUDA-aware Cray MPI was made. However, no speedup was observed because RDMA is only efficient for smaller message sizes due to its limited bandwidth despite a lower communication latency. The strong scaling limit for our code is around 512 to 1024 nodes, which corresponds to the point when communication overhead dominates the entire simulation. Increasing node count beyond this point results in worse performance.

\subsection{Amphiphilic polymer self-assembly}

A simulation of spontaneous vesicle formation using 134,217,728 particles is performed to further demonstrate the capability of our new DPD program to study complex fluids. An aqueous surfactant solution system consisting of water and an amphiphilic triblock copolymer were modeled. The particles representing water, hydrophilic block and hydrophobic block are denoted as S, A, B, respectively. The triblock copolymer has a chain configuration of BBBAABBB. The system was initialized to be a random distribution of water particles and polymer molecules. Bounce-forward boundary condition was applied instead of the periodic boundary condition. Beads within a molecule are tied together using a harmonic potential $F=K(r-r_0)$, where the coefficients were chosen as $r_0=0.38$ and $K=80$. The DPD parameters were chosen as $\rho = 5.0$, $\sigma=3.0$, $\gamma=4.5$, $k_B T = 1.0$, and $\delta t=0.01$. The pairwise repulsive force magnitudes $a_{ij}$ are shown below:
\begin{align*}
\left(
\begin{array}{ccccc}
   &   A &   B &   S \\
A  &  15 & 120 &  15 \\
B  & 120 &  15 & 120 \\
S  &  15 & 120 &  15
\end{array}
\right)
\end{align*}

\begin{figure}[htbp]
  \centering
  \subfigure[t=10]{
    \includegraphics[width=2.25in]{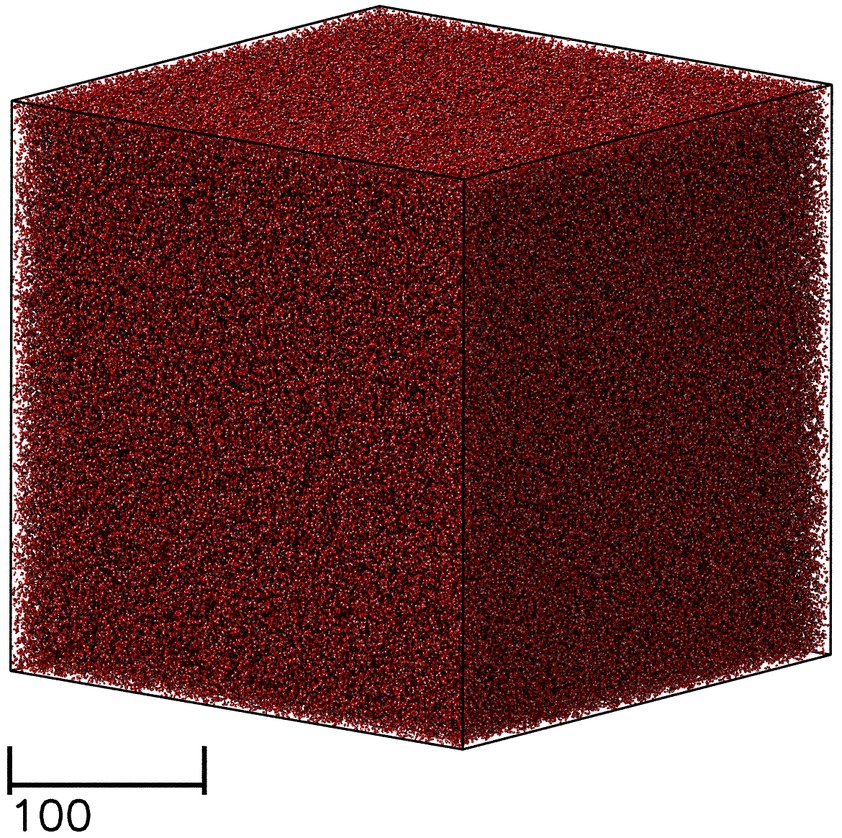}
    \label{fig:128m_overview_a}
  }
  \subfigure[t=1000]{
    \includegraphics[width=2.25in]{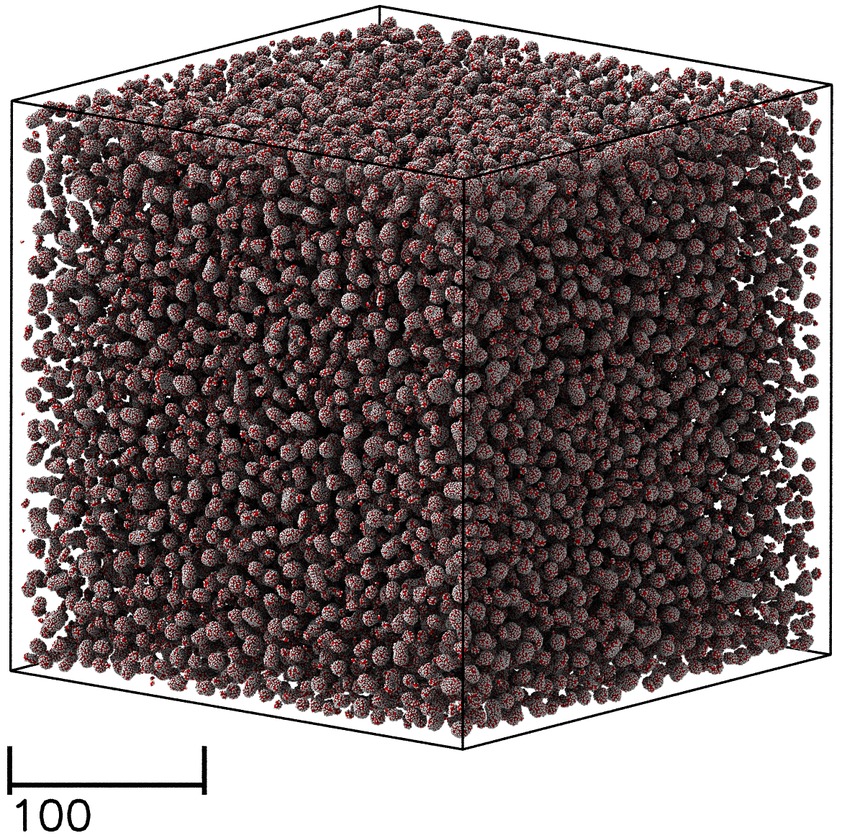}
    \label{fig:128m_overview_b}
  }
  \subfigure[t=15000]{
    \includegraphics[width=2.25in]{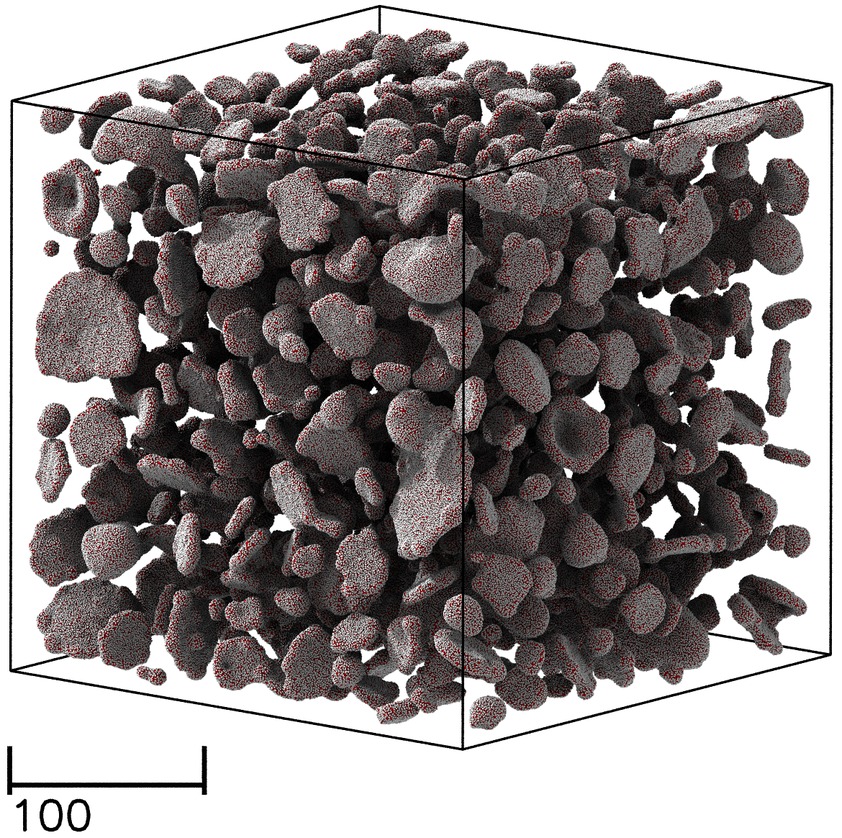}
    \label{fig:128m_overview_c}
  }
  \subfigure[t=55000]{
    \includegraphics[width=2.25in]{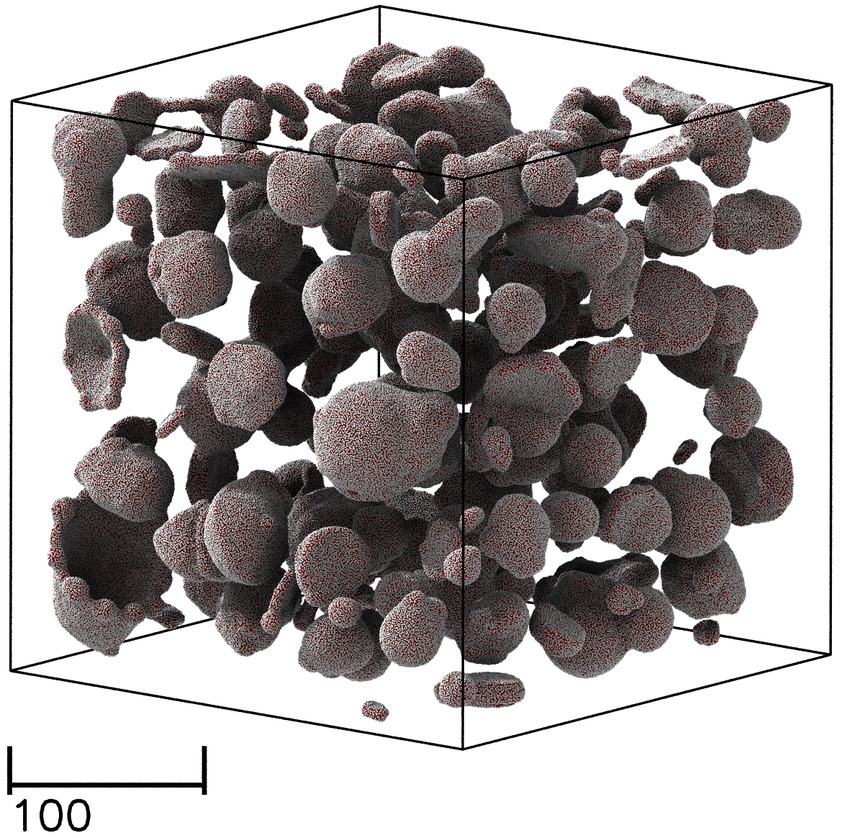}
    \label{fig:128m_overview_d}
  }
  \subfigure[t=124000]{
    \includegraphics[width=2.25in]{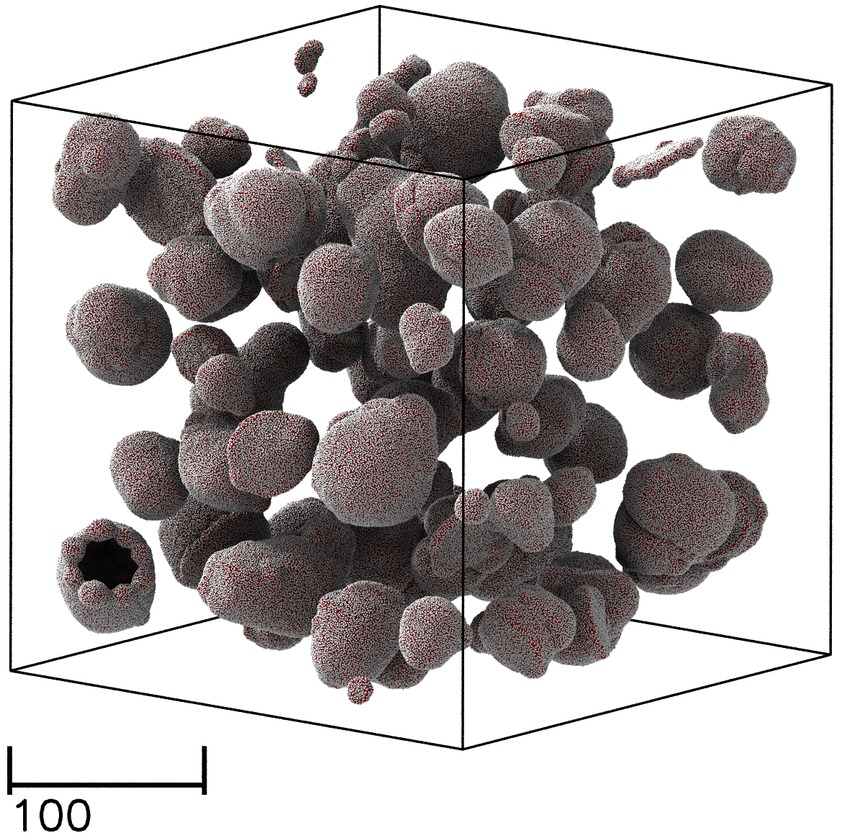}
    \label{fig:128m_overview_e}
  }
  \caption{A variety of vesicles and membrane-like structures were grown from the 128 million particle simulation. The BBBAABBB triblock copolymer was initially randomly distributed in a cubic system of size $299.4\times299.4\times299.4$. The concentration of the surfactant is 10\%. Scale bars are in reduced DPD units.}
  \label{fig:128m_overview}
\end{figure}

\begin{figure}[htbp]
  \centering
  \subfigure[]{ \includegraphics[width=3in]{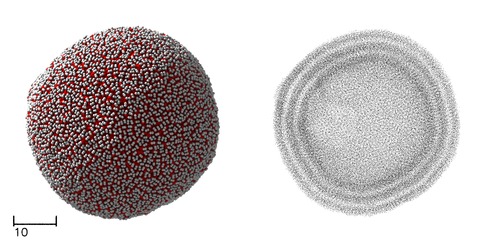} \label{fig:vesicle_32} }
  \subfigure[]{ \includegraphics[width=3in]{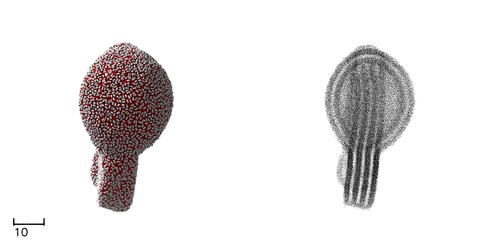} \label{fig:vesicle_65} }
  \subfigure[]{ \includegraphics[width=3in]{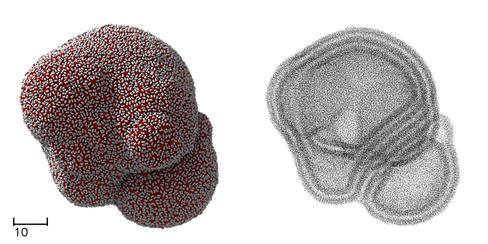} \label{fig:vesicle_10} }
  \subfigure[]{ \includegraphics[width=3in]{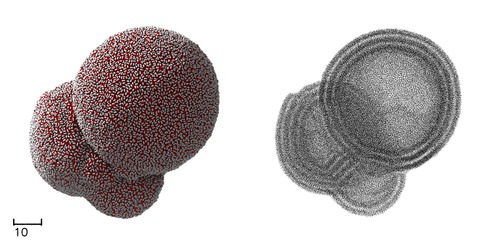} \label{fig:vesicle_41} }
  \subfigure[]{ \includegraphics[width=3in]{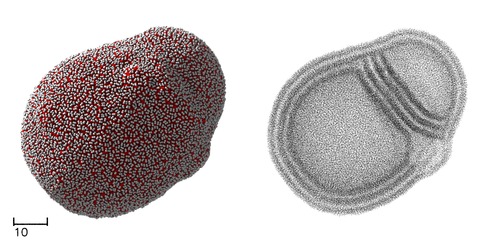} \label{fig:vesicle_76} }
  \subfigure[]{ \includegraphics[width=3in]{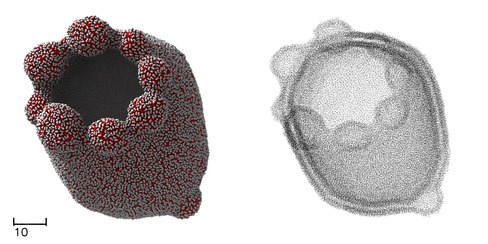} \label{fig:vesicle_69} }
  \subfigure[]{
    \includegraphics[height=1.75in]{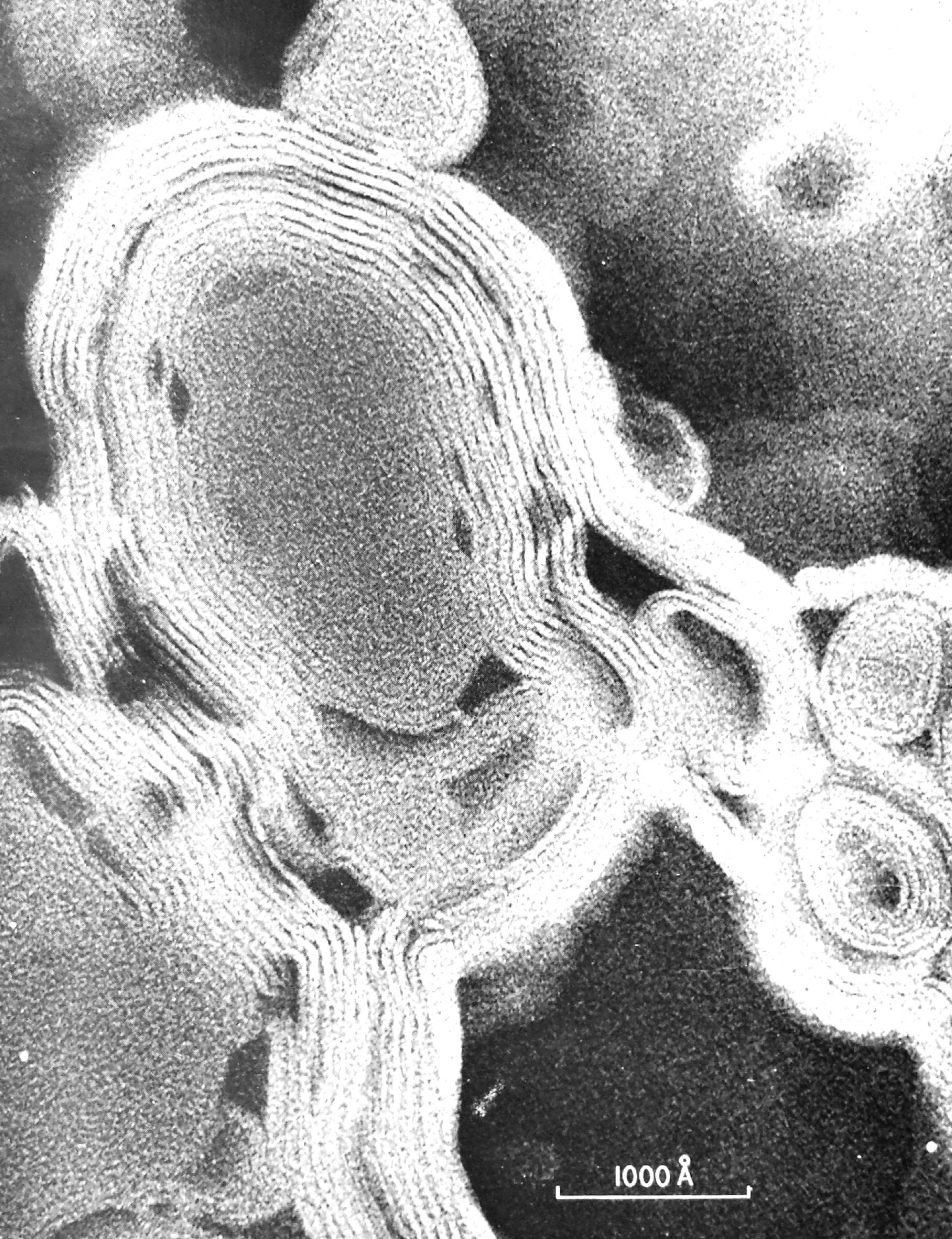}
    \label{fig:onion_vesicle_a}
  }
  \subfigure[]{
    \includegraphics[height=1.75in]{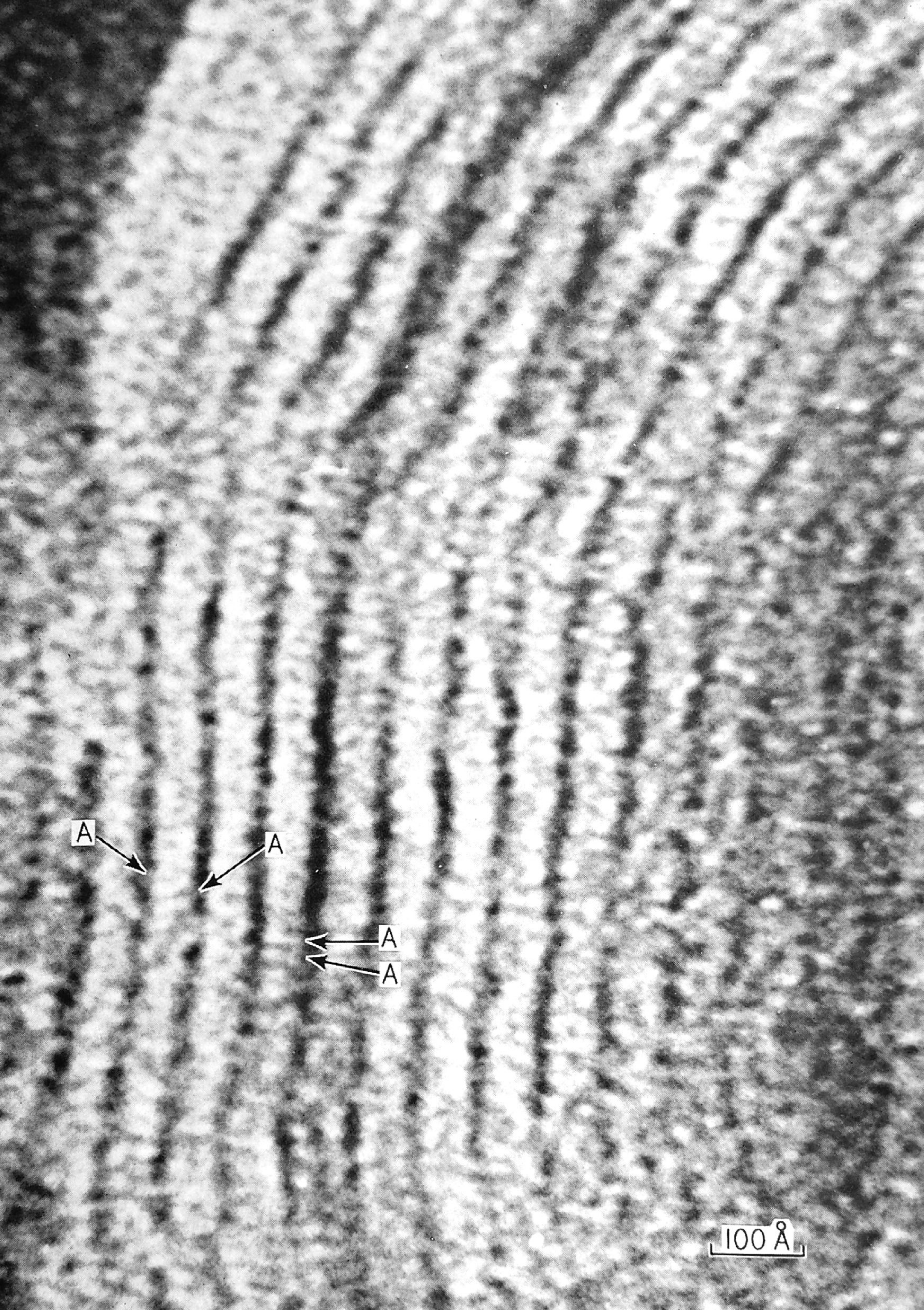}
    \label{fig:onion_vesicle_b}
  }
  \caption{The multi-walled vesicle shown in \subref{fig:vesicle_32}-\subref{fig:vesicle_69} exhibit notable structural similarity to that observed by Bandham \textit{et al} \cite{Bangham1964660} in \subref{fig:onion_vesicle_a} and \subref{fig:onion_vesicle_b} (reprinted with permission from Elsevier). The hydrophilic and hydrophobic particles are rendered in white and red, respectively, in the colored images; and in white and black, respectively, in the transmission microscopy-style images. Scale bars are in reduced DPD units.}
  \label{fig:vesicle_growth}
\end{figure}

\begin{figure}[htbp]
  \centering
  \subfigure[t=90]    { \includegraphics[width=2in]{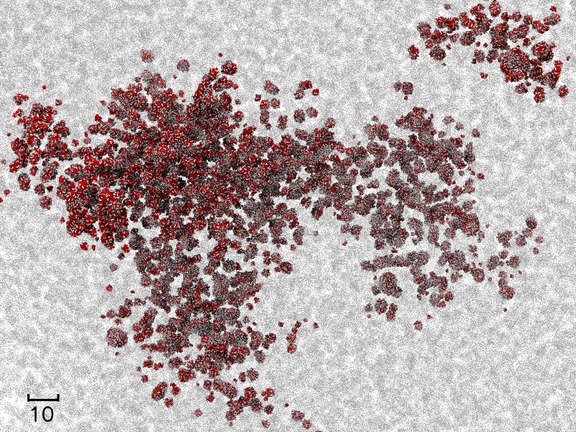} \label{fig:vesicle_32_proc_a} }
  \subfigure[t=3100]  { \includegraphics[width=2in]{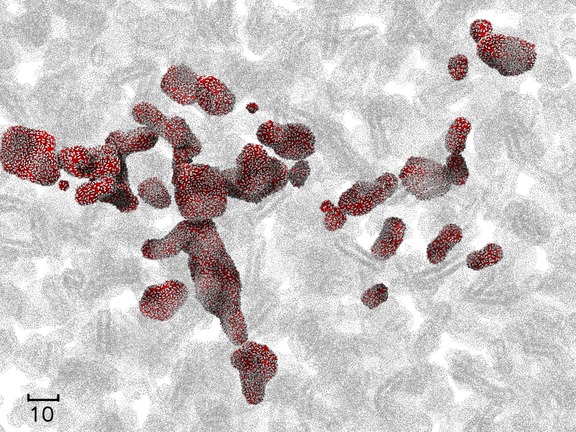} \label{fig:vesicle_32_proc_b} }
  \subfigure[t=27100] { \includegraphics[width=2in]{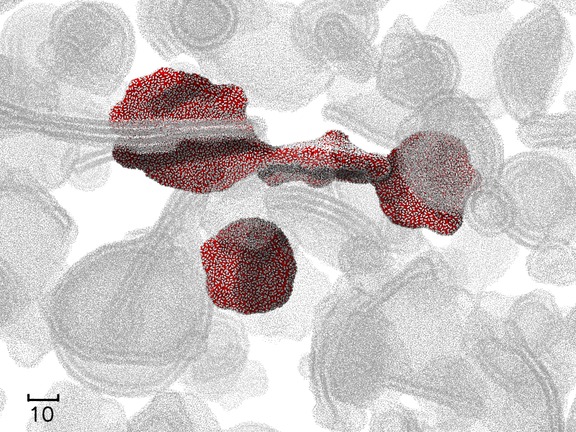} \label{fig:vesicle_32_proc_c} }
  \subfigure[t=36100] { \includegraphics[width=2in]{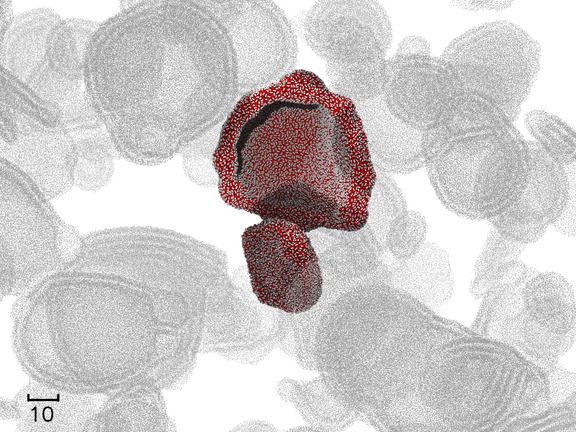} \label{fig:vesicle_32_proc_d} }
  \subfigure[t=111100]{ \includegraphics[width=2in]{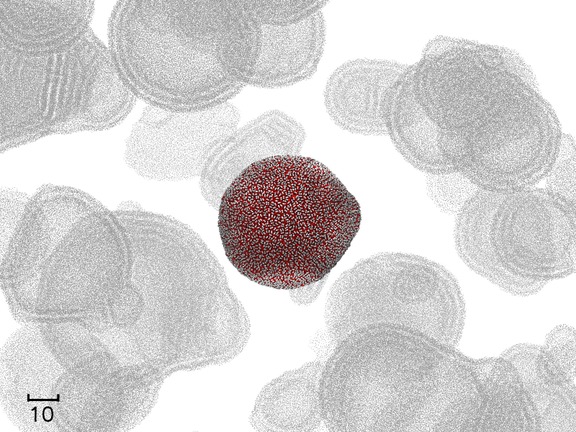} \label{fig:vesicle_32_proc_e} }
  \caption{The vesicle shown in Figure~\ref{fig:vesicle_32} is formed through the fusion and spontaneous wrapping of membrane-like structures. Polymer molecules that eventually became part of the vesicle are highlighted, while irrelevant ones are drawn as shadows. Scale bars are in reduced DPD units.}
  \label{fig:vesicle_dynamics}
\end{figure}

The simulation took 54 hours to complete on 1024 nodes of TITAN. A total of 12,400,000 time steps, or a time span of 124,000 in reduced DPD uints, were sampled. The time evolution of the system as illustrated in Figure~\ref{fig:128m_overview} proves that a mixture of membranes, simple vesicles and multi-compartment vesicles with a remarkable morphological variation were grown from the initially random distribution. The structure of the vesicles as visualized by a transmission electron microscopy-style rendering technique exhibits many similarities to those of the onion-like multilamellar multicompartment vesicles observed by Bangham \textit{et al} \cite{Bangham1964660} as shown in Figure~\ref{fig:vesicle_growth}. Figure~\ref{fig:vesicle_dynamics} demonstrates that the single-compartment vesicle in Figure~\ref{fig:vesicle_32} is actually formed through the fusion and spontaneous wrapping of membrane-like structures.

\section{Conclusion}
\label{sec:conslusion}

We presented a complete design of the DPD simulation code on the CUDA GPUs. The program is derived from LAMMPS and it achieves substantial speedup over the original CPU version. Novel algorithms are used in all parts of the program including force evaluation, data layout and numerics. An atomics-free, deterministic algorithm for constructing the neighbor list is proposed. We have also been able to address locality and neighbor list building in a tightly-coupled framework. Numerical optimization targeting random number generation and functional form evaluation were proposed to fully utilize the computing power of the GPU. The code executes on large-scale supercomputers with near-optimal weak scaling efficiency, while strong-scaling efficiency is compromised by the GPU-CPU data transfer overhead. A large-scale simulation of spontaneous vesicle formation demonstrated the practicality of our code. Future work includes optimizing communication latency through overlapping computation with communication as well as load balancing for non-trivial domain geometry. The code will be contributed as a LAMMPS user module that is freely available to the general public.

\clearpage

\FloatBarrier



\begin{table}[htbp]
\caption{Achieved bandwidth of the data marshaling and bandwidth-sensitive kernels. The benchmarks were done on a Kepler K20 GPU with a peak memory bandwidth of 208 GB/s.}
\centering
\begin{tabular}{c c c c}
\hline \hline
Kernel & Read (GB/s) & Write (GB/s) & Aggregate (GB/s) \\
\hline
Interleave            & 79.59 & 79.62 & 159.21 \\
Deinterleave          & 72.21 & 70.80 & 143.01 \\
NeighborListJoin      & 62.66 & 57.83 & 120.49 \\
NeighborListTranspose & 78.03 & 79.06 & 157.09 \\
MergeXVT              & 95.78 & 54.66 & 150.44 \\
\hline
\end{tabular}
\label{table:interleave_bandwidth}
\end{table}

\begin{table}[htbp]
\caption{Performance comparison of the neighbor list builders using warp ballot or atomics increment.}
\centering
\begin{tabular}{c c c c}
\hline \hline
\multirow{2}{*}{system config} & \multicolumn{2}{c}{time(ms)} & \multirow{2}{*}{speedup} \\
                        & ballot        & atomic       &  \\
\hline
$\rho=3,  n=64k$  &  0.76 &   1.55 & 2.02 \\
$\rho=3,  n=512k$ &  5.28 &  10.77 & 2.04 \\
$\rho=6,  n=64k$  &  1.31 &   2.92 & 2.24 \\
$\rho=6,  n=512k$ &  9.89 &  21.72 & 2.20 \\
$\rho=50, n=64k$  & 14.58 &  30.15 & 2.07 \\
$\rho=50, n=512k$ & 90.28 & 196.26 & 2.17 \\
\hline
\end{tabular}
\label{table:neigh_builder}
\end{table}

\begin{table}[htbp]
\caption{Comparison between our custom double-precision math routines (prefixed with {\em fast}) and the CUDA native ones. This microbenchmark is done through a chain of dependent statements according to Ref ~\cite{microbenchmark}.}
\centering
\begin{tabular}{r c c c c}
\hline \hline
Function & Instruction\# & Conditional Branch & Latency(cycles) & Max Rel. Error \\
\hline
fastcospi   & 23  & 0  &   90  &  $1.10\times 10^{-10}$ \\
cospi       & 54  & 3  &  420  &  1 ULP \\
fastlog     & 47  & 0  &  380  &  $4.21\times 10^{-12}$ \\
log         & 89  & 5  &  603  &  1 ULP \\
fastpow     & 79  & 0  &  494  & 11 ULP \\
pow         & 258 & 18 & 1982  &  2 ULP \\
\hline
\end{tabular}
\label{table:microbench}
\end{table}

\clearpage

\nolinenumbers

\section*{Acknowledgement}
\label{sec:acknow}

This work was supported by the new Department of Energy Collaboratory on Mathematics for Mesoscopic Modeling of Materials (CM4).
Y.H.T. acknowledges discussion and technical support from Michael L. Parks, Steve Plimpton and Christian Trott at Sandia National Laboratories.
Simulations were carried out at the Oak Ridge Leadership Computing Facility through the Innovative and Novel Computational Impact on Theory and Experiment program at Oak Ridge National Laboratory under project BIP017.

\section*{References}
\label{sec:references}
\bibliography{v5-arXiv}
\bibliographystyle{unsrt-abbrev}

\end{document}